\documentclass[fleqn,usenatbib]{rasti}

\usepackage{newtxtext,newtxmath}
\usepackage[T1]{fontenc}

\DeclareRobustCommand{\VAN}[3]{#2}
\let\VANthebibliography\thebibliography
\def\thebibliography{\DeclareRobustCommand{\VAN}[3]{##3}\VANthebibliography}

\usepackage{multirow, enumitem, enumerate}

\usepackage{amsmath, amsthm, booktabs, threeparttable}
\usepackage{graphicx, color, xcolor}
\usepackage{comment}

\newtheorem{proposition}{Proposition}
\newtheorem{definition}{Definition}
\definecolor{darkgreen}{rgb}{0.0, 0.4, 0.0}

\def\gridline#1{\vskip6pt\hbox to\hsize{#1}\vskip6pt}

\def\fig#1#2#3{\hfill\vbox{\parskip=0pt\hsize=#2
\includegraphics[width=#2]{#1}\vskip2pt\vtop{\centering
\footnotesize
\hsize=#2
#3\vskip1pt
}}\hfill}
\title[Concordance]{Extended Bayesian Estimation for In-flight Calibration of Space-based Instruments}

\author[]{Soham Das,$^{1}$
Yang Chen,$^{1}$, 
David A. van Dyk$^{2}$\thanks{Contact e-mail: dvandyk@imperial.ac.uk},
Herman L. Marshall$^{3}$,
Vinay L. Kashyap$^{4}$, 
\newauthor
and 
Michael Smith$^{5}$ \\
$^{1}$University of Michigan, Ann Arbor, MI, USA\\
$^{2}$Statistics Section, Department of Mathematics, Imperial College London, 180 Queen’s Gate, London, SW7 2AZ, UK\\
$^{3}$Kavli Institute for Astrophysics and Space Research, Massachusetts Institute of Technology, Cambridge, MA, USA\\
$^{4}$ Center for Astrophysics | Harvard and Smithsonian, 60 Garden Street, Cambridge, MA 02138, USA\\
$^{5}$ Telespazio UK for European Space Agency, ESAC, Camino Bajo del Castillo, s/n, Villanueva de la Cañada, E-28692 Madrid, Spain\\
}

\date{Accepted XXX. Received YYY; in original form ZZZ}

\pubyear{\the\year{}}

\begin{document}
\label{firstpage}
\pagerange{\pageref{firstpage}--\pageref{lastpage}}
\maketitle

\begin{abstract}
The ongoing calibration of space-based instruments is essential to ensure accurate measurements of physical quantities such as photon flux. The goal is to obtain a unified photon-flux measurement of a single astronomical source using data obtained with multiple instruments. Calibration involves in-flight adjustments to address discrepancies arising from instrument-specific variations. Taking a statistical approach, we extend a previously proposed additive log-normal hierarchical model with half-variance correction based on a multiplicative physical model for photon counts. The newly proposed Bayesian hierarchical model can leverage domain-specific priors to account for correlated passbands and integrate information across different instruments, sources, and passbands effectively. The model is fitted using Markov chain Monte Carlo and can provide uncertainty quantification for instrument- and source-specific parameter estimation. Moreover, we obtain valid inferences for other structural parameters of the physics model. We use data derived from \textit{Chandra} and XMM-\textit{Newton} X-ray observatories to illustrate the proposed inference procedure.
\end{abstract}

% Include between one and six keywords.
\begin{keywords}
Bayesian hierarchical model -- high energy spectral model -- posterior inference% -- shrinkage estimate
\end{keywords}

%%%%%%%%%%%%%%%%%%%%%%%%%%%%%%%%%%%%%%%%%%%%%%%%%%
%%%%%%%%%%%%%%%%%%%%%%%%%%%%%%%%%%%%%%%%%%%%%%%%%%

%%%%%%%%%%%%%%%%% BODY OF PAPER %%%%%%%%%%%%%%%%%%

\section{Introduction}\label{sec:introduction}

We use space-based instruments for measurement of high-energy signatures of astronomical source, including, most fundamentally, the energy of photons and the sky coordinates of their sources. The accuracy of the measurements may be compromised by harsh launch conditions and prolonged orbital operation. The discrepancies between operational and pre-launch measurement accuracy may arise from instrument-specific variations, intrinsic properties of the astronomical objects, and inaccuracies or shifts in the receiver module's readings. Additional factors such as gas leakage, filter degradation, component failure, and contamination buildup can further affect performance \citep{marshall2021concordance}. Calibration of the instruments is required to avoid systematic errors in photon flux measurements. We focus on improving calibration with a better estimate of the instrumental \textit{effective area}, particularly to model the uncertainty associated with shifts in the flux measurements. The effective area is considered as an equivalent area over which the instrument effectively collects photons, and it is usually smaller than the instrument's geometric area.

Calibration is typically performed in two phases: first in a controlled laboratory environment and second through in-flight calibration. Ground-based measurements are insufficient for this purpose, as they are taken under environmental conditions that differ significantly from those encountered in orbit \citep{weisskopf2002chandra}. It is also insufficient to compare observations of different sources obtained with a single instrument, because this does not account for instrument-specific biases, see \citet{lee2011accounting}. In this article, we focus exclusively on in-flight calibration methods that leverage comparisons of data from multiple photon sources using multiple instruments. Developing a procedure for obtaining reliable adjustments from data collected across multiple objects and instruments is known as the calibration concordance problem \citep{chen2019calibration}. Two notable studies, \citet{chen2019calibration} and \citet{marshall2021concordance}, have made progress in this area by applying Bayesian methods to model the effective area of high-energy astronomical instruments using statistical shrinkage estimates. Although these models include energy passband-specific factors, they ignore interactions between photon sources and passbands. Indeed, there is no established general calibration procedure for in-flight instrument calibration that accounts for spectral distortions. This work aims to establish statistically principled calibration standards using data from multiple instruments with multiple correlated bandpasses observing the same set of sources, thereby improving upon the empirical procedures traditionally employed by the astronomy community. More precisely, in this paper, we account for spectral distortions in effective area estimates, seeking to (partially) resolve the calibration concordance problem in the high-energy spectral models.

We introduce a Bayesian hierarchical model that incorporates passband-wise correlation and a parametric spectral model for source fluxes, neither of which has been considered in the literature. Due to the large number of unknown parameters relative to the available number of observations, the resulting model is highly flexible but only weakly identifiable, meaning some parameters are not uniquely estimable without additional constraints. Hence, we further constrain the model with a parametric spectrum for the source fluxes. Taking a Bayesian perspective, we use the resulting posterior distribution to estimate the model parameters and their uncertainty, to evaluate model performance, and to assess calibration. This requires us to obtain a Monte Carlo sample from the posterior distribution. Given the complexity of this distribution, we use Hamiltonian Monte Carlo (HMC) with a No-U-Turn Sampler (NUTS); see \citet{neal1993probabilistic} and \citet{hoffman2014no} for details of HMC and NUTS. 
A notable advantage of the proposed model, given the complexity and (in)feasibility of obtaining the desired astronomical observations, is that we do not require all passbands to be observed with all instruments for all sources. We show theoretically that the available data can be used to obtain valid estimates of the effective area and source fluxes. Thus, our model provides reliable estimates and uncertainties for instrument- and passband-specific calibration corrections and for the spectral parameters;
all of which can be used for subsequent measurement corrections and instrument adjustments.

The remainder of this paper is organized as follows. In Section~\ref{sec:data_measurement_framework}, we present a detailed overview of the instrument calibration data and standard formulations for necessary physical quantities. Section~\ref{sec:methodology} describes the proposed methodology, including the motivation for our approach and the specification of the models. In this section, we introduce three models that differ in the degree of constraint imposed through their prior structures. Section~\ref{sec:model_evaluation} presents theoretical properties of the models and describes the information criterion used to select the most appropriate specification. In Section~\ref{sec:results}, we illustrate the application of our concordance models to the calibration data obtained from the \textit{Chandra} and XMM-\textit{Newton} X-ray observatories, and then compare the performance of the competing models. Using the selected model, we obtain (i) corrected effective areas for all instrument-passband combinations, (ii) estimates of the structural parameters of the spectral model, and (iii) reliable estimates of source fluxes. We also show how our calibration model can be used to examine changes in the effective areas of XMM-Newton's \texttt{MOS1} and \texttt{MOS2} instruments over time. Finally, Section~\ref{sec:conclusion} summarizes our contributions and outlines future directions for the methodological development of calibration concordance.

%%%%%%%%%%%%%%%%%%%%%%%%%%%%%%%%%%%%%%%%%%%%%%%%%%%%%%%%%%%%%%%%%%%%%%%%%%%%%%%%%%%%%%%%%%%%%%%%%%%%%%%%%%%%%%%%%%%%%%%%%%%%%%%%%%%%%%%%%%%%%%%%%%%%%%

\section{Data and Measurement Framework}\label{sec:data_measurement_framework}

We motivate calibration concordance by reviewing the fundamental quantities governing X-ray measurements. In particular, we describe how the source flux, energy passband, and instrument effective area jointly determine the transformation of photon spectra into the detector counts used for calibration. We then describe the data analyzed in this article.

%%%%%%%%%%%%%%%%%%%%%%%%%%%%%%%%%%%%%%%%%%%%%%%%%%

\subsection{Effective Area and Source Flux}\label{sec:fundamental_quantity}

Throughout this paper, we use  $i$, $s$, and $p$ to index instrument, source, and passband, respectively, and refer to each using its enumerated label. The photon energy $E$ in our study ranges from $0.15$ to $ 10.0$ keV and is divided into ten passbands. Each energy passband interval is denoted by $\mathcal{E}_p$ where $p=1, \ldots, P$ and $P=10$. These passbands are ordinal variables, with lower passbands corresponding to lower energy levels. 

Let $n_E(\eta_s)$ be the photon spectrum for source $s$ in units of photons per unit area 
{
at energy $E$, where 
$\eta_s$ represents the vector of spectral parameters for source $s$. 
The spectral shape is given by 
\begin{equation}\label{eqn:spectral_shape}
    q(E; {\tilde{\eta}_s}) = 
    \frac{{E} \ n_E(\eta_s)}
    {{E_0}\ n_s} 
\end{equation}
as a function of $\tilde{\eta}_s{\equiv (n_s, \eta_s) }$, {where $E_0=1$ keV}  and $n_s$ is the overall normalization factor with the same units as $n_E$.}

The total flux (in units of erg/cm$^2$/s) of source $s$ in energy passband $\mathcal{E}_p$ is denoted by
\begin{equation}
\label{eqn:flux}
F_{sp} = 
\int_{\mathcal{E}_p} 
{E}  \
n_E({\eta}_s) \ {\rm d}E = n_s {E_0}\int_{\mathcal{E}_p} q(E; {{\eta}_s)} \ {\rm d}E.
\end{equation}
{The definition of source flux used in this paper differs from that in \citet{marshall2021concordance}. (This is due to the difference in the methods used to compute the fluxes, see Section~\ref{subsec:observed_c}.) In this paper fluxes have units erg/cm2/s, resulting in an extra factor of energy (erg) in the integral over the bandpass in Equation~\eqref{eqn:flux} as compared to \citet{marshall2021concordance}.}

Although the objective of space-based X-ray telescopes is {often} to measure the photon flux $q(E;\tilde{\eta}_s)$, in this work, we focus on the calibration of the instruments and bandpasses by analyzing the concordance of flux measurements observed by each instrument. We cannot directly compare the observed flux with its true value, since the true photon flux is never known. Therefore, we need to introduce the concept of \textit{effective area}, which quantifies the photon collection efficiency of the instruments, and ultimately enables the comparison of observed flux values across instruments and sources.

We formulate the effective area of instrument $i$ in passband $p$ as $A_{ip}  \alpha_{i}(E)$, where $\alpha_{i}(E)$ represents the known shape of the effective area curve for instument $i$. We assume that the $\alpha_{i}(E)$ is known but the $A_{ip}$ is not, reflecting the fact that the relative effective area across a narrow band of energies is generally better known than its absolute value. This is the same assumption made in \citet{chen2019calibration} and \citet{marshall2021concordance}.

We consider an absorbed power-law spectrum in each of the ten passbands in Definition~\ref{def:spectralmodel}. 
\begin{definition}[X-ray Spectral Model] \label{def:spectralmodel}
The {spectral shape} for source $s$ %$n_E(\theta_s)$ 
(in \textrm{ph~cm$^{-2}$~s$^{-1}$~keV$^{-1}$}) is given by
\begin{equation}\label{eqn:spectral_model}
q(E;{\tilde{\eta}_s}) = 
 {\left(\frac{E}{E_0}\right)^{1-\tilde{\Gamma}_s} }
 \exp\Big\{-N_H \sigma(E) + \gamma_s [\, \log(E/{E_0})\,]^2\Big\},
\end{equation}
{where $\tilde{\eta}_s = ({n_s}, \tilde{\Gamma}_s, \gamma_s)$,} $N_{\rm H}$ is the {(assumed known)} interstellar medium (ISM) column density in atoms/cm$^2$, and $\sigma(E)$ is the cross-section of the ISM in cm$^2$. 
This results in a band-wise log-scale flux from the above spectral model with parameters $\theta_s = (a, b, c)$ given by 
\begin{equation}
\label{eqn:spectrum_model_log}
    g_p(\theta_s) = a + \log \int_{E\in \mathcal{E}_p} E^{1-b + c \log E} e^{-N_{\rm H} \sigma(E)}{\rm d}E, 
\end{equation}
{where $a=\log(n_s) +\tilde{\Gamma}_s\log(E_0)$, $b=\tilde{\Gamma}_s$, and $c=\gamma_s$. The dependency of $(a, b, c)$ on index $s$ is surpressed to simplify notation.}
\end{definition}

In practice, $\sigma(E)$ is calculated approximately within each energy band, as we detail in Definition~\ref{def:model_3}. In Section~\ref{subsec:observed_c}, we describe a crude procedure to derive physically meaningful ranges for $a, b,$ and $c$, which are used {as starting values} in the subsequent Bayesian model. 
This procedure uses the calibration data in a minimal manner in an effort to avoid double use of the data. 

%%%%%%%%%%%%%%%%%%%%%%%%%%%%%%%%%%%%%%%%%%%%%%%%%

\subsection{Multiplicative Model for Expected Counts}\label{subsec:multiplicativemodel_expected}

Suppose that the passband $\mathcal{E}_p$ of instrument $i$ is composed of $\mathcal{K}_{ip}$ photon-counting detector channels. The \textit{expected} photon count in channel $k$ of passband $\mathcal{E}_p$ is related to the true photon energy $E$ via the product of photon spectrum, effective area, and exposure time. We denote this expected count by 
\begin{equation}
C_{ispk} = t_{is}  \int_{\mathcal{E}_p} A_{ip}\alpha_{i}(E) n_E(\eta_s) \Phi_{ik}(E) {\rm d}E,
\end{equation}
where $t_{is}$ is the exposure time. The instrument response function $\Phi_{ik}(E)$ tabulates the probability that a photon of energy $E$ is recorded in detector channel $k$, with
\begin{equation}
\sum_p\sum_{k\in\mathcal{K}_{ip}} \Phi_{ik}(E) = 1.
\end{equation}
Plugging in the expressions for $n_E(\eta_s)$ and $F_{sp}$ given in Equations \eqref{eqn:spectral_shape} and \eqref{eqn:flux}, we obtain $C_{ispk} =\hat{T}_{ispk} \,A_{ip} \,F_{sp}\,/\,E_0 $,
where 
\begin{equation}\label{eqn:estimate_t}
\hat{T}_{ispk}(\eta_s) =
t_{is}\frac
{\int_{\mathcal{E}_p} \alpha_{i}(E)\,{\frac{E_0}{E}}q(E; \eta_s)\Phi_{ik}(E)\,{\rm d}E}
{\int_{\mathcal{E}_p} q(E; \eta_s) \,{\rm d}E}
\end{equation} 
can be calculated numerically given the effective area curve $\alpha_{i}(E)$ and the parameter $\theta_s$. Aggregating over the passbands, we obtain 
\begin{equation}\label{eqn:mean_formula}
C_{isp}
=\sum_{k\in\mathcal{K}_{ip}}
\ \hat{T}_{ispk}\ A_{ip}\ F_{sp} / {E_0} 
= \hat{T}_{isp}\ A_{ip}\ F_{sp}/ {E_0} .
\end{equation}
\noindent
For simplicity, we refer to 
$A_{ip}$ as the `effective area'.

%%%%%%%%%%%%%%%%%%%%%%%%%%%%%%%%%%%%%%%%%%%%%%%%%%

\subsection{Observed Photon Fluxes}\label{subsec:observed_c}

Due to complications of data recording, instead of directly using observed photon counts, we follow a standard protocol to derive photon-count summaries, $c_{isp}$, an equivalent observed version of the expected photon count $C_{isp}$.  
In this section, we describe how we compute observed fluxes (which gives $c_{isp}$). This material is not required for the main results and is included for readers interested in the technical details.

The photon flux $\hat{f}_{isp}$ is calculated {from Equation~\eqref{eqn:flux} and Definition~\ref{def:spectralmodel}} 
through an absorbed power-law spectrum with known column density $N_H$ and unknown slope $\tilde{\Gamma}_{is}$,

\noindent
where $n_{0,is}$ {(corresponding to $n_s$ in Equation~\eqref{eqn:flux})} is a known normalization factor at reference energy $E_0=1keV$ and the flux units are $10^{-11}$ erg/cm$^2$/s.  Except for setting $N_{\rm H}$ to externally determined values specific to each object, we do not attempt to tie the spectral parameters fitted to each band. For each instrument and GTI (Good Time Interval), the absorbed flux in each band is computed using the model \eqref{eqn:spectrum_model_log} by numerical integration using the fitted values. We expect the values of ${\hat f_{isp}}$ to depend only weakly on the estimates of $\tilde{\Gamma}_{is}$: small deviations would not result in significant changes. Not all instruments are sensitive to all the passbands we consider. Thus, some of the passband-specific fits were unreliable due to insufficient signal.

Calibration scientists provide default estimates $a_{ip}$ and measurement uncertainties $\tau_{ip}$ for the effective areas $A_{ip}$. In Section~\ref{sec:methodology}, we use $a_{ip}$ and $\tau_{ip}$ to formulate prior distributions for the effective areas in our Bayesian model. Using the $a_{ip}$, the passband-wise photon counts are calculated (based on Equation~\eqref{eqn:mean_formula}) as
\begin{equation}\label{eqn:count_isp}
c_{isp} = \sum_{k\in{\cal K}_{ip}} c_{ispk}
\equiv \, \hat{T}_{isp} \, a_{ip} \,\hat{f}_{isp} / {E_0} .
\end{equation}
This gives the observed photon fluxes, i.e., noisy measurements of $C_{isp}$ that we use as data.

%%%%%%%%%%%%%%%%%%%%%%%%%%%%%%%%%%%%%%%%%%%%%%%%%%

\subsection{Data Description}\label{subsec:data_description}

\begin{figure}
  %\centering
  \includegraphics[scale=0.45]{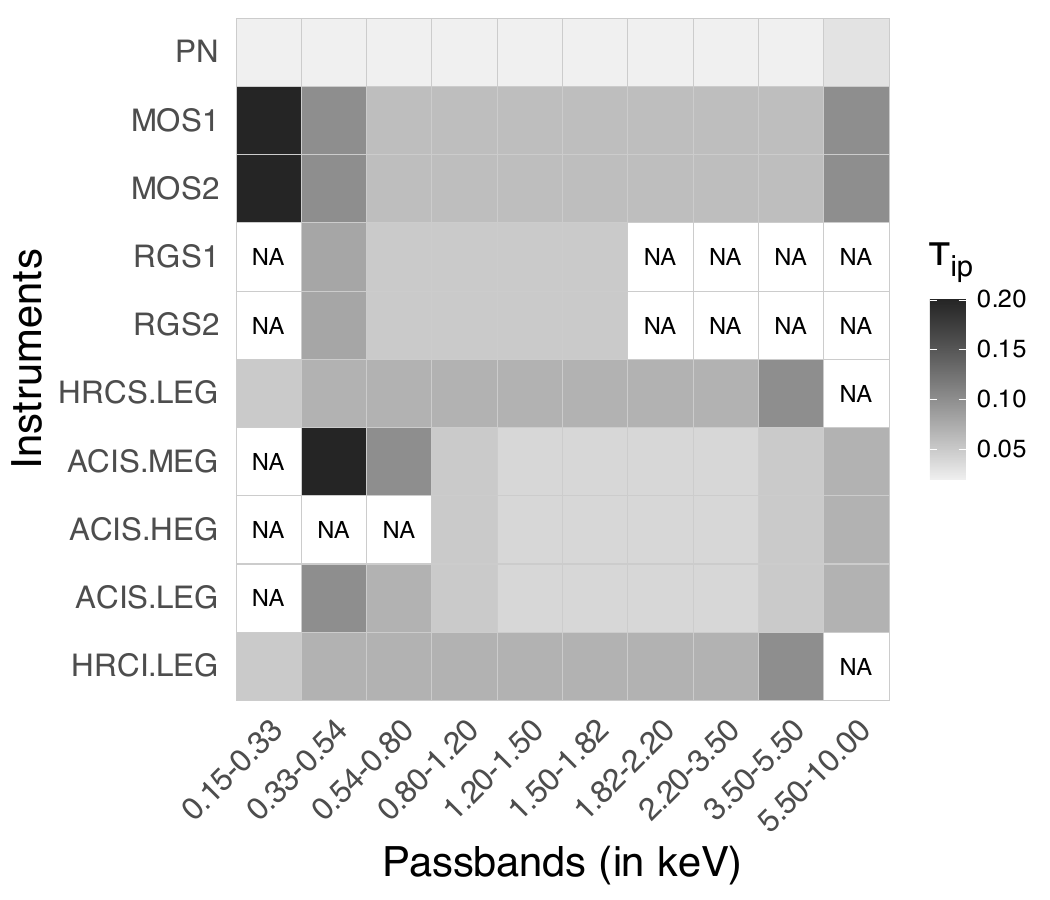}
  \caption{Heatmap of prior standard deviation $\tau_{ip}$ for relative error of effective area across all ten instruments and ten passbands. The shade of each cell is proportional to the values of the standard deviations, provided by IACHEC scientists. The missing values are represented by white cells labeled `NA'. Due to absence of prior variance information, all observations corresponding to these cases have been discarded, and effective area parameters corresponding to these cells cannot not estimated.}
  \label{fig:heatmap_tau}
\end{figure}
The data analyzed in this study are similar to those used in \citet{marshall2021concordance}. They were observed using ten instruments, each on board the \textit{Chandra} or the XMM-\textit{Newton} X-ray observatory and are  selected because they are sufficiently distinct to warrant cross-calibration using multiple source-instrument combinations. A subset of {45} XCAL sources was selected for detailed analysis; see Figure \ref{fig:heatmap_tau} for an instrument list. Each source was observed by at least two instruments (and at least once by each of \textit{Chandra} and XMM-\textit{Newton}). The sources are from four AGNs, specifically, from the quasars/blazars: 3C 273, Mk 590, PKS 2155-304, and 1H 1426$+$428.
We refer to these AGNs as `targets'. Table~\ref{tab:target_count} lists the number of sources considered for each of the AGN targets and the names of 10 of the 45 sources appear in Table~\ref{tab:source_counts}. The interstellar medium (ISM) column density values, $N_H$, adopted in our analysis for four targets are 1.79, 1.55, 1.61, and 3.07 (in units of atoms/cm$^2$) for 3C 273, 1H 1426$+$428, Mk 590, and PKS 2155$-$304, respectively.

Due to flux and spectral variability, only strictly simultaneous observations are used, resulting in a unique GTI (good time interval) for each data set, which we use as a source label. XMM allows simultaneous data collection in all five of its instruments (\texttt{PN}, \texttt{MOS1}, \texttt{MOS2}, \texttt{RGS1}, \texttt{RGS2}), as does \textit{Chandra} in the {HEG} and {MEG} datasets. Thus, these data share the same time intervals.  For datasets which include XMM, there are up to four XMM instruments, and for datasets which include \textit{Chandra}, there are either one or two instruments with the same interval (i.e., the same GTIs).

For each source-instrument pair, detected photons in the energy range $[0.15, 10.00]$ keV were tabulated into passbands to obtain an observed spectrum. We use the passbands defined in \citet{marshall2021concordance}, with the exception that the $1.2-1.8$ keV band was subdivided into two bands, $1.2-1.5$ keV and $1.5-1.8$ keV. 
All \textit{Chandra} measurements employed a grating, and only the dispersed spectra were used in the analysis. 
The resolutions (i.e., accuracy of photon energy measurement) and the {effective areas} of the various instruments differ significantly. Thus, initial corrections are applied to the data using {default calibration quantities} 
that are known to be only approximately correct.
\begin{table}
\centering
\caption{Number of sources associated to each of the four AGN targets in the dataset.}
\label{tab:target_count}
\begin{tabular}{ccccc}
\toprule
&\multicolumn{4}{c}{AGN Targets}\\
\cmidrule{2-5}
& 3C273 & H1426+428 & Mkn590 & PKS2155-304 \\ 
\midrule
Source count & 18 & 2 & 3 & 22\\
\bottomrule
\end{tabular}
\end{table}

We remove poorly determined source measurements from the data by fitting a simple model, which we describe below, to determine initial estimates of the spectral parameters $(a,b,c)$ and then remove data points where $|g_p-g_p(\hat{\theta})| > 0.56$, indicating a flux that deviates from the model by a factor of 1.75 or more. The flux in Equation~\eqref{eqn:spectrum_model_log}, given the spectral parameters $\theta=(a,b,c)$, can be approximated by
\begin{align}
 \label{eqn:flux_model}
   g_p(a,b,c) & = a + K_p + \alpha_p b + \beta_p c + \gamma_p b c,
   \end{align}
where the $p$th band is defined by energy limits $[E_{1p},E_{2p}]$, $K_p  = \xi_p + d_p - N_{\rm H} \sigma_p$, $\alpha_p  = -\xi_p $, $\beta_p  = 0.014 + 0.132 \xi_p + \xi_p^2$, $\gamma_p  = -0.05 \xi_p${, $\xi_p \equiv \log E_{0p}$, $ E_{0p} \equiv (E_{2p} + E_{1p})/2$, $d_p \equiv \log (E_{2p} - E_{1p})$ }, and $\sigma_p$ is an emperical approximation of $\sigma(E)$ in the passband $p$ given right after Definition~\ref{def:model_3}.
The fitting formula in \eqref{eqn:flux_model}  provides approximations of $g_p$ that are accurate to about 0.01 with one or two exceptions at low energies due to the vagaries of poorly known interstellar gas abundances. The process of obtaining Equation~\eqref{eqn:flux_model} involves some algebra, physics intuition, and trial and error, as the absorption model is not an analytic function but a lookup table with discontinuities which generally appear at the bandpass boundaries. The range of guaranteed accuracy is for $1.5 < b < 3$ and $-0.3 < c< 0.5$.  The values of $a$ are expected to be in the range $[0.1,10]$ in units of keV~cm$^{-2}$~s$^{-1}$. The fluxes in the blazar XCAL data file are in units of $10^{-11}$ erg~cm$^{-2}$~s$^{-1}$, which differs from the model units by a factor of $1.6{\times}10^{-9}$ erg~keV$^{-1}$.
\begin{table}
\centering
\caption{Joint instrument observation frequencies of 10 sources and 10 passbands for 10 out of 45 sources. The ten passbands are represented by their indices.}
\label{tab:source_counts}
\begin{tabular}{ccccccccccc}
\toprule
\multicolumn{1}{c}{\multirow{2}{*}{Sources}} & \multicolumn{10}{c}{Passband Indices}\\
\cmidrule{2-11}
 & {1} & {2} & {3} & {4} & {5} & {6} & {7} & {8} & {9} & {10}\\
\midrule
\texttt{2462}  & 4 & 6 & 6 & 6 & 6 & 6 & 4 & 4 & 4 & 3 \\
\texttt{2463}  & 3 & 6 & 6 & 7 & 7 & 7 & 5 & 5 & 5 & 5 \\
\texttt{2464A} & 3 & 6 & 6 & 6 & 5 & 5 & 4 & 4 & 4 & 4 \\
\texttt{2464B} & 3 & 6 & 6 & 6 & 6 & 6 & 4 & 4 & 4 & 4 \\
\texttt{5169A} & 2 & 4 & 5 & 6 & 6 & 6 & 3 & 4 & 4 & 4 \\
\texttt{5169B} & 2 & 5 & 6 & 7 & 7 & 7 & 5 & 5 & 5 & 4 \\
\texttt{5170A} & 3 & 6 & 6 & 6 & 5 & 5 & 4 & 4 & 4 & 4 \\
\texttt{5170B} & 2 & 6 & 6 & 6 & 6 & 6 & 4 & 4 & 4 & 4 \\
\texttt{4430}  & 3 & 6 & 6 & 6 & 6 & 6 & 4 & 4 & 4 & 4 \\
\texttt{4431}  & 3 & 6 & 6 & 6 & 6 & 6 & 4 & 4 & 4 & 4 \\
\bottomrule
\end{tabular}
\end{table}

In summary, photon spectra from a total of $45$ sources are measured simultaneously by at least two of the 10 instruments, in a subset of 10 energy passbands. These are organized into a photon count data cube with elements $c_{isp}$, where $i=1,\ldots, 10$ indexes instruments, $s=1,\ldots,45$ indexes  GTI sources, and $p=1,\ldots, 10$ indexes the passbands. This yields $4500$ potential combinations of instruments, sources, and passbands. Each combination has either zero or one observation, which means that we either observe a particular combination or it is missing. 

\begin{figure}
  %\centering
  \includegraphics[scale=0.4]{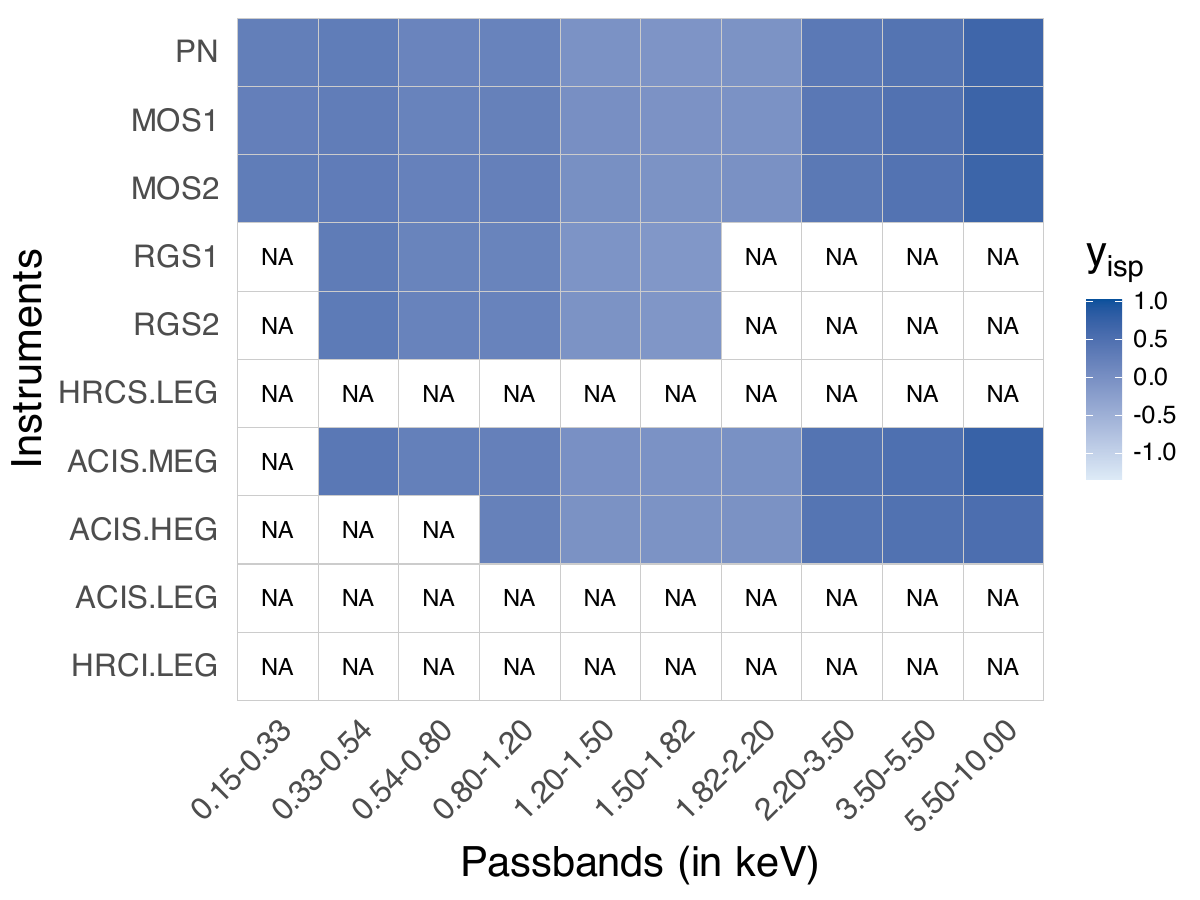}
  \caption{Heatmap of observed log-transformed photon count $y_{isp}=\log c_{isp}$ (as defined in Equation~\eqref{eqn:count_isp}) for source \texttt{2463}, for ten instruments and ten passbands. The color shades of cells are proportional to corresponding observed quantity $y_{isp}$, while the missing values are denoted as `NA'.}
    \label{fig:photon_heatmap}
\end{figure}

Out of the 4500 possible entries in the data cube, many observations are missing for different reasons. For example, since our aim is to develop a robust calibration framework, we exclusively focus on high-flux values. Thus, zero-count observations which typically arise from weak photon flux or instrumental limitations have been excluded, reducing the data to 2252 observed entries. Figure~\ref{fig:photon_heatmap} illustrates the log-transformed observed photon count for the source \texttt{2463}, where a large number of the observations are not available (marked in `NA'). 

We also excluded observations that lacked measurement uncertainties for $c_{isp}$, denoted as $\nu_{isp}$, which reduced the data cube to 2155 available observations. Additionally, we removed observations related to specific instrument-passband combinations for which expert-provided uncertainties, denoted as $\tau_{ip}$, for the effective areas were not available. The rationale for such deletion is discussed in Section \ref{sec:log_transformed_model}. As illustrated in Figure~\ref{fig:heatmap_tau}, uncertainties are missing for 17 instrument-passband combinations. After these filtering steps, the final dataset comprises of 2098 observations, for which we assume the missingness is independent of the observations from all 10 instruments, 45 photon sources, and 10 passbands. Table~\ref{tab:source_counts} and \ref{tab:instrument_counts} present two-way summary tables of the observed frequencies for each source-passband combination and instrument-passband combination respectively. 

%%%%%%%%%%%%%%%%%%%%%%%%%%%%%%%%%%%%%%%%%%%%%%%%%%%%%%%%%%%%%%%%%%%%%%%%%%%%%%%%%%%%%%%%%%%%%%%%%%%%%%%%%%%%%%%%%%%%%%%%%%%%%%%%%%%%%%%%%%%%%%%%%%%

\section{Methodology}\label{sec:methodology}

In this section, we outline our model framework and justify its assumptions on dependency structures and prior distributions. Following calibration concordance literature, we begin with a multiplicative model for the expected photon counts and then use a log transformation to obtain an additive model for which Gaussian assumptions are straightforward to justify. Three models are considered, based on different prior structures, in order of increasing complexity; the last model is the one we newly propose in this paper, which accounts for the spectral model and various dependency structures in effective areas and source fluxes.

\begin{table}
\centering
\setlength{\tabcolsep}{5pt}
\caption{Joint frequencies of 10 instruments and 10 passbands. The ten passbands are represented by their indices.}
\label{tab:instrument_counts}
\begin{tabular}{lcccccccccc}
\toprule
\multicolumn{1}{c}{\multirow{2}{*}{\footnotesize Instruments}} & \multicolumn{10}{c}{Passband Indices}\\
\cmidrule{2-11}
            & {1} & {2} & {3} & {4} & {5} & {6} & {7} & {8} & {9} & {10}\\
\midrule
\texttt{PN}        & 39 & 43 & 43 & 43 & 43 & 43 & 42 & 43 & 43 & 43 \\
\texttt{MOS1}      & 33 & 41 & 41 & 41 & 40 & 40 & 40 & 41 & 41 & 41 \\
\texttt{MOS2}      & 36 & 38 & 38 & 38 & 37 & 37 & 37 & 38 & 38 & 38 \\
\texttt{RGS1}      &  0 & 45 & 45 & 45 & 42 & 42 &  0 &  0 &  0 &  0 \\
\texttt{RGS2}      &  0 & 45 & 45 & 45 & 39 & 39 &  0 &  0 &  0 &  0 \\
\texttt{HRCS.LEG}  &  8 &  8 &  8 &  8 &  8 &  8 &  7 &  8 &  5 &  0 \\
\texttt{ACIS.MEG}  &  0 &  6 & 15 & 16 & 16 & 16 & 16 & 16 & 16 & 10 \\
\texttt{ACIS.HEG}  &  0 &  0 &  0 & 16 & 14 & 14 & 16 & 17 & 17 & 17 \\
\texttt{ACIS.LEG}  &  0 & 17 & 17 & 17 & 17 & 17 & 17 & 17 & 17 & 16 \\
\texttt{HRCI.LEG}  &  1 &  1 &  1 &  1 &  1 &  1 &  1 &  1 &  1 &  0 \\
\bottomrule
\end{tabular}
\end{table}

%%%%%%%%%%%%%%%%%%%%%%%%%%%%%%%%%%%%%%%%%%%%%%%%%

\subsection{Logarithm Additive Model in Literature}\label{sec:log_transformed_model}

In Section \ref{sec:data_measurement_framework}, we formulated the calibration concordance problem into a parametric estimation setup, which we can rely on to obtain inferences of instrument-specific effective area parameters $A_{ip}$ and source-specific flux parameters $F_{sp}$ from the passband-wise aggregated observed photon counts $c_{isp}$ defined in Section~\ref{subsec:observed_c}. The multiplicative model given in Equation~\eqref{eqn:mean_formula} in Section~\ref{subsec:multiplicativemodel_expected} yields a naive ratio estimate. More precisely, given prior estimates $a_{ip}$ for the effective area $A_{ip}$, a simple way to estimate $F_{sp}$ is by computing the ratio $\hat{F}_{sp}=\displaystyle\frac{{E_0} \sum_i c_{isp}}{\sum_i a_{ip}\hat{T}_{isp}}$. However, this approach is challenging since this ratio estimator is neither unbiased nor consistent. Instead, \citet{chen2019calibration} and \citet{marshall2021concordance} model the observed photon counts $c_{isp}$ by introducing a multiplicative log-normal error $\xi_{isp}$, i.e., 
\begin{equation}\label{eqn:multiplicative_model}
    c_{isp}= \hat{T}_{isp}\, A_{ip}\, F_{sp}\, \xi_{isp}\, \exp(-\sigma_i^2/2) {/E_0},
\end{equation}
where $\log \xi_{isp}\sim \mathcal{N}(0,\sigma_i^2)$; the adjustment $\exp(-\sigma_i^2/2)$ ensures the expected value $C_{isp} =\mathbb{E}(c_{isp})$,
since for log-normal distribution, $\mathbb{E}[X]=\exp(\sigma^2/2)$ when $\log X\sim \mathcal{N}(0,\sigma^2)$. Note that the error distribution is assumed to depend only on the instrument for simplicity.

As given in \citet{chen2019calibration} and \citet{marshall2021concordance}, we rewrite the equation \eqref{eqn:multiplicative_model} after taking the logarithm on both sides as
\begin{equation}\label{eqn:log_model}
    y_{isp}= -\frac{\sigma_i^2}{2} + B_{ip} + G_{sp} + \varepsilon_{isp},
\end{equation}
where $B_{ip}=\log A_{ip}-\log a_{ip}$, $G_{sp}=\log F_{sp}$ and $\varepsilon_{isp}\sim\mathcal{N}(0,\sigma_i^2)$. Similarly, we subtract previously estimated $\log a_{ip}$ and $\log \hat{T}_{isp}$ to define ${y}_{isp}=\log c_{isp} - \log\hat{T}_{isp} - \log a_{ip}-${$\log E_0$}, assuming that all observed photon counts are positive (as mentioned in Section \ref{subsec:data_description}). The transformed log-photon count $y_{isp}$ is the response variable for our log-transformed model. Equation \eqref{eqn:log_model} serves as our working model equation for instrument calibration, with $B_{ip}$, $G_{sp}$ and $\sigma_i^2$ as its model parameters.

The presence of the half-variance correction term of $-\frac{1}{2}\sigma_i^2$ in Equation \eqref{eqn:log_model} distinguishes it from being a simple linear regression equation. As a result, the posterior mean and variance estimates for each parameter are dependent on each other. Any bias in the posterior variance estimates can therefore propagate to the posterior of location estimates, leading to inflated residual errors. This interdependence creates the need for a carefully specified model that adequately captures variability arising from measurement error, passband-wise correlation, instrument, and source–specific effects. Given the resulting model structure, closed-form expressions for posterior summaries are unavailable. Thus, we need to use sampling-based inference methods, where we draw samples from the joint posterior distribution of parameters using the Markov chain Monte Carlo algorithms. 
We can rewrite Equation~\eqref{eqn:log_model} using a vectorized form for notational convenience. We  
define
\begin{align*}
     \boldsymbol{y}_{is}&=(y_{is1},\ldots,y_{isP}),\\ \boldsymbol{B}_{i}&=(B_{i1},\ldots,B_{iP}), \\
     \boldsymbol{G}_{s}&=(G_{s1},\ldots,G_{sP}). 
\end{align*}
The resulting vectorized model equation is
\begin{equation}\label{eqn:log_vector_model}
    \boldsymbol{y}_{is} = -0.5\sigma_i^2\boldsymbol{1}_P + \boldsymbol{B}_i + \boldsymbol{G}_s + \boldsymbol{\varepsilon}_{is},
\end{equation}
where $\boldsymbol{\varepsilon}_{is}\sim\mathcal{N}(\boldsymbol{0}, \sigma_i^2I_P)$; $\boldsymbol{1}_P$ denotes a vector of ones of length $P$ and $I_P$ is the identity matrix of size $P$. Throughout the rest of this paper, boldface characters denote vectors, and normal font characters denote scalars and matrices. We use vector notation as needed to simplify the expressions.

With 2098 observations and 542 parameters, the model is only weakly identifiable. Thus, we employ hierarchical modeling to pool information and obtain stable estimates. By constraining parameters that share common indices through appropriate prior distributions, we create inter-dependency within the parameters at the posterior level. As a result, the posterior distributions of the parameters with common indices share information. For example, the posterior distribution of each effective-area relative error parameter $B_{ip}$ is directly dependent on observations that are measured by the same instrument-passband combination and its own prior distribution via data likelihood and prior likelihood respectively. It can indirectly depend on effective area parameters from the other instruments and passbands through their posterior distributions. Creating such dependencies through hierarchical structure reduces posterior estimation uncertainty and improves stability. It also restores parameter identifiability under mild conditions; more discussion on this is given in Section \ref{sec:identifiability}. This property can be further extended to predict certain missing observations which can be used to check the predictive performance of our model. A more detailed discussion is given in Appendix~\ref{appx:pvalue}.

We define Model 1, originally proposed in \citet{chen2019calibration} with a simpler data structure, as a baseline calibration model. In this model, the effective-area error vector $\boldsymbol{B}_i$ is assigned a Gaussian prior with no correlation across passbands, the log-flux vector $\boldsymbol{G}_s$ follows a non-informative prior, and the instrument-specific variance $\sigma_i^2$ is assigned a conjugate inverse-gamma prior with parameters $(3, 2\nu_i^2)$. We further assume the prior covariance matrix for $\boldsymbol{B}_i$ to be diagonal, with diagonal entries $\tau_{ip}^2$ provided by calibration scientists (Figure~\ref{fig:heatmap_tau}). Photon count uncertainties $\nu_{isp}$ are originally specified by observers for each individual observation. We define an instrument-specific summary $\nu_i$ as the harmonic mean across sources and passbands, given by $\nu_i^{-2}=\frac{1}{SP}\sum_s\sum_p\nu^{-2}_{isp}$. The prior hyperparameters of prior of $\sigma_i^2$ are chosen to be Inverse-Gamma$(3,2\nu_i^2)$ so that the prior mean of $\sigma_i^2$ matches the corresponding uncertainty value $\nu_i^2$. Under conjugate prior assignment, the resulting (conditional) posterior estimates of $B_{ip}$ and $G_{sp}$ admit a shrinkage estimator formulation, which ensures stable estimation, see Section \ref{sec:identifiability}.
\begin{definition}[\textbf{Model 1}: Baseline Calibration Model without Correlation Structures in \citet{chen2019calibration}]\label{model_1}
This model is given by Equation \eqref{eqn:log_vector_model} with the priors for model parameters, effective area error $\boldsymbol{B}_i \sim \mathcal{N}(0,\Gamma_i)$ where $\Gamma_i = \text{diag}(\boldsymbol{\tau}_i^2)$ is a diagonal matrix, log-flux {$\boldsymbol{G}_s \sim \text{non-informative prior}$}, and $\sigma_i^2\sim \text{Inverse-Gamma(3, 2$\nu_i^2$)}$.
\end{definition}

If the number of observations in passband $p$ of instrument $i$ is low, the posterior distribution of $B_{ip}$ is more reliant on its prior distribution. As a result, the posterior estimates of $B_{ip}$ from low-frequency class can be sensitive to prior mis-specification. Since the primary goal of this study is to systematically model the data measurement uncertainties, a mis-specification in the prior uncertainty estimate can introduce bias in the posterior distribution of model parameters. So we remove all observations $y_{isp}$ that do not have prior measurement uncertainty $\nu_{isp}$. Additionally, imbalance in the instrument-wise frequencies of observations can create bias in the posterior estimates of the parameters from low-frequency class. This behavior is reflected in the \texttt{HRCI.LEG} instrument HRCI.LEG, where the posterior estimates of $B_{ip}$ have a higher spread with a small portion lying outside the tolerance boundary, see Figure \ref{fig:appx_instrument_plot}-(h).

In this situation, small mis-specification in the prior can lead to a biased posterior estimate for the effective area. This is one of the reasons why the prior hyperparameter specification is crucial for posterior analysis. To improve the quality of the estimation and avoid bias, all observations corresponding to settings where prior hyperparameters are not available are discarded from our study.

\begin{figure}
  \centering
  \includegraphics[scale=0.45]{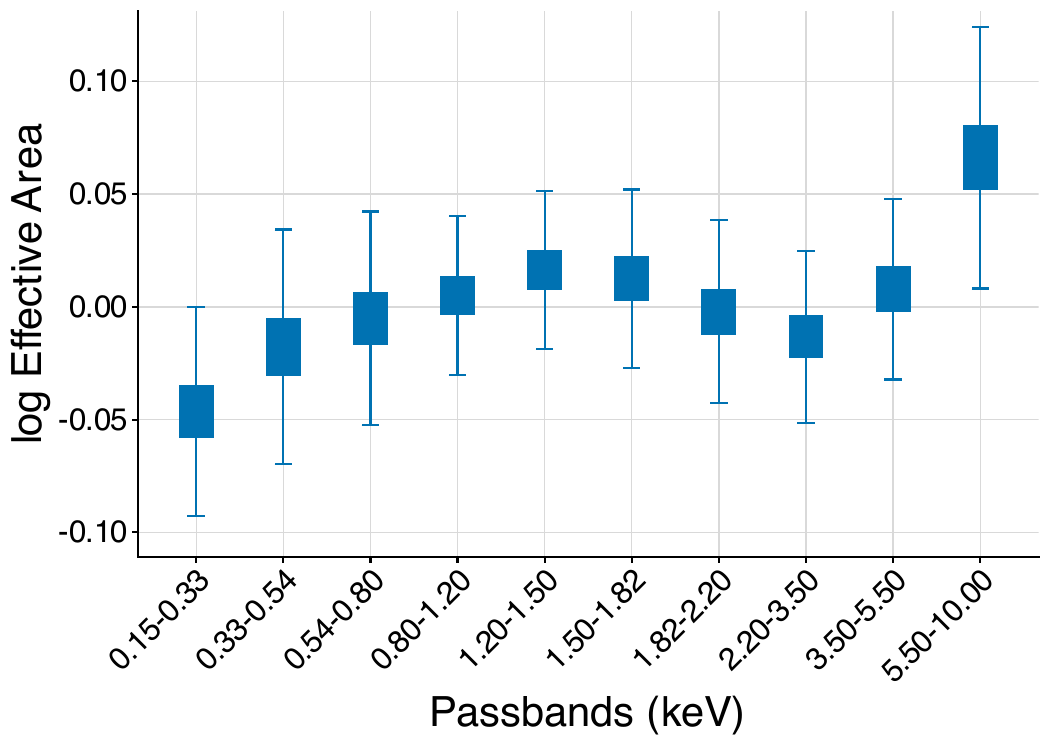}
  \caption{Boxplots of posterior samples of the effective area relative error $B_{ip}$ for instrument \texttt{MOS1} across all passbands, obtained from Model 1 in Definition~\ref{model_1}. The rectangle of each boxplot spans the inter-quartile range or IQR (50\% posterior credible interval), while the whiskers denote $1.5$ times the IQR of posterior estimates of $B_{ip}$ in each passband. If the posterior estimates lie within $\pm$0.20, %then it implies that 
  the relative error of estimated effective area are within $20\%$.}
\label{fig:instrument_sample_plot}
\end{figure}

Figure \ref{fig:instrument_sample_plot} shows boxplots of the posterior distributions of $B_{ip}$ for $i=$ \texttt{MOS1} for all passbands from Model 1 in Definition~\ref{model_1}. Here, the rectangular box is the inter-quartile range and the full span of a boxplot is $1.5$ times the inter-quartile range. This represents our baseline model. 
Across all passbands, the inter-quartile ranges lie within the $\pm 0.2$ pre-specified tolerance band for the logarithm of effective areas by calibration scientists, corresponding to a 20\% tolerance bound for the effective area $A_{ip}$ on the original scale.

%%%%%%%%%%%%%%%%%%%%%%%%%%%%%%%%%%%%%%%%%%%%%%%%%%

\subsection{Correlation Across Passbands}\label{sec:correlation_passbands}

Model 2 (given in Definition~\ref{def:model_2}) is proposed in \citet{marshall2021concordance}, where the correlations of effective-area errors across passbands are introduced for each instrument. The rationale for modeling passband-wise correlation is to validate the astrophysical phenomenon that the errors from a single instrument are inherently correlated across energy bands. 

More precisely, we assume $\mathbf{B}_i\sim\mathcal{N}(0,\Gamma_i)$, where $\Gamma_i$ is not diagonal as in Model 1 (Definition~\ref{model_1}). The prior correlations (the $R_i$ matrix in Definitions~\ref{def:model_2}, \ref{def:model_3} and \ref{def:model_3_ext}) are calculated by calibration scientists using Monte Carlo integration; see \citet{drake2007monte,kashyap2008handle,xu2014fully} for details. Since the Monte Carlo methods used to derive correlations are experimental, we do not use the same methods when computing the mean and variance parameters.  For the prior mean and variance (denoted as $a_{ip}$ and $\tau_{ip}^2$ respectively), we rely on the values provided by calibration scientists, as detailed in IACHEC meetings from 2017–19 \citep{madsen2019summary, madsen2020summary} and elaborated in \citet{marshall2021concordance}. The $\tau_{ip}$ values that we use in this paper are presented as a heatmap in Figure~\ref{fig:heatmap_tau}. 

\begin{definition}[\textbf{Model 2}: Calibration Model with Correlation among Passbands in \citet{marshall2021concordance}] This model is given by Equation \eqref{eqn:log_vector_model} with the priors for model parameters specified as: effective area relative error $\boldsymbol{B}_i \sim \mathcal{N}(0,\Gamma_i)$ where $\Gamma_i = \text{diag}(\boldsymbol{\tau}_i)R_i\text{diag}(\boldsymbol{\tau}_i)$ is a known covariance matrix with correlation matrix $R_i$, log-flux {$\boldsymbol{G}_s \sim \text{non-informative prior}$}, and $\sigma_i^2\sim \text{Inverse-Gamma(3, 2$\nu_i^2$)}$.
\label{def:model_2}
\end{definition}

%%%%%%%%%%%%%%%%%%%%%%%%%%%%%%%%%%%%%%%%%%%%%%%%%%

\subsection{Newly Proposed Hierarchical Model}\label{sec:hierarchical_model}
In this section, we describe our newly proposed hierarchical model, extending the existing work described in Section~\ref{sec:log_transformed_model} by considering source flux models and structural dependencies of hyperparameters.  {We adopt the power-law formula for the photon flux in Equation~\eqref{eqn:spectrum_model_log} as a prior structure for the log-flux $\boldsymbol{G}_{s}$ in Model 3 given in Definition~\ref{def:model_3}.}

\begin{definition}[\label{def:model_3}\textbf{Model 3}: Calibration Model with Correlation among Passbands and a Spectral Model for Source Fluxes.] This model is given by Equation \eqref{eqn:log_vector_model} with the priors for model parameters specified as: relative error of effective area, $\boldsymbol{B}_i \sim \mathcal{N}(0,\Gamma_i)$ where $\Gamma_i = \text{diag}(\boldsymbol{\tau}_i)R_i\text{diag}(\boldsymbol{\tau}_i)$ is a covariance matrix with known correlation matrix $R_i$, $\sigma_i^2\sim \text{Inverse-Gamma(3, 2$\nu_i^2$)}$, and log-flux, {$\boldsymbol{G}_s\sim \mathcal{N}_P(\boldsymbol{g}({\theta}), \Lambda_s$)}, with $\boldsymbol{g}(\theta)=(g_1(\theta),\ldots,g_P(\theta))$, where $g_p(\theta)$ is given in Equation~\eqref{eqn:spectrum_model_log}.
Priors for $(a, b, c)$ are assigned based on expert recommendations: $b\sim\text{Uniform}(1.5, 3)$, $c\sim\text{Uniform}(-0.3, 0.5)$ and $a$ follows a non-informative, flat prior.
\end{definition}

{To reduce computation time we approximate $\sigma(E)$ in Equation~\eqref{eqn:spectrum_model_log}. Defining passband $p$, i.e., $\mathcal{E}_p$, by energy limits $[E_{1p},E_{2p}]$, we approximate its $\sigma(E) = (E/\tilde{E}_p)^{-\zeta_p}$}  as $\sigma_p = (E_{0p}/\tilde E_p)^{-\zeta_p}$, where $E_{0p}=[(E_{2p} + E_{1p})/2]$, $\tilde{E}_p = 0.192$ and $\zeta_p = 2.83$ for $p=1$ and $p=2$, and $\tilde E_p = 0.185$ keV and $\zeta_p = 2.4$ for all other passbands $p$. Numerically, this approximation does not significantly change parameter estimation or posterior inference.

Due to significant variations in photon emission over time, we treat one AGN target as different sources when it is observed at different GTI. Each of the 45 photon sources is observed from one of the four nuclei targets. Names of these targets and the number of sources corresponding to each of the targets is given in Table \ref{tab:target_count}. This indicates that the true flux spectrum originating from the same target should have a similar shape. Moreover, the scale of the true flux also depends on its source. By using target- and source-dependent hyperparameters in the prior specification for the log-flux parameter $\boldsymbol{G}_{s}$, we create multiple sub-models and variants within Model 3, which are defined in Definition~\ref{def:model_3_ext}. The variants of Model 3 are classified into four sub-models (denoted Model 3A, 3B, 3C, 3D) and or obtained by varying the prior covariance structure of $\boldsymbol{G}_s$. The sub-models are created to cover all four combinations of involving heterogeneity and correlation in the prior covariance of $\boldsymbol{G}_s$. Further variants of Model 3 are considered following the structural choices of the prior means of $\boldsymbol{G}_s$. Intuitively, separate estimation of source- and target-dependent hyperparameters may yield a better fit than the simplest version of Model~3 since it allows for more flexibility.
\begin{table*}
    \centering
    \caption{Overview of the prior specifications across the three models. Model 1 does not have passabnd-wise correlation in $\boldsymbol{B}_i$, but Model 2 includes correlation between passbands of same instrument through correlation matrix $R_i$. In Model 3, we assume informative Gaussian prior with spectral parameters as mean and variance. For simplicity, the sub-model distinctions within Model 3 are not shown.}
    \begin{tabular}{c|c|c|c}
    \toprule
        \multirow{2}{*}{} & \multicolumn{3}{c}{Priors}\\
        \cmidrule{2-4}
    
        Models & effective-area rel. error & log-flux & error variance\\
        \midrule
            
        \multirow{2}{*}{Model 1} & $\boldsymbol{B}_i \sim \mathcal{N}(0,\Gamma_i)$ & \multirow{2}{*}{$\boldsymbol{G}_s \sim \text{non-informative prior}$} & 
        \multirow{2}{*}{$\sigma_i^2\sim$
        \text{IG}$(3,2\nu_i^2)$}
        \\
        & $\Gamma_i = \text{diag}(\boldsymbol{\tau}_i^2)$ & & \\
        \cmidrule{1-4}
        
        \multirow{2}{*}{Model 2} & $\boldsymbol{B}_i \sim \mathcal{N}(0,\Gamma_i)$ & \multirow{2}{*}{$\boldsymbol{G}_s \sim \text{non-informative prior}$} & 
        \multirow{2}{*}{$\sigma_i^2\sim$
        \text{IG}$(3,2\nu_i^2)$}
        \\
        & $\Gamma_i = \text{diag}(\boldsymbol{\tau}_i)R_i\text{diag}(\boldsymbol{\tau}_i)$ & & \\
        \cmidrule{1-4}

        \multirow{4}{*}{Model 3} & 
        & $\boldsymbol{G}_s \sim N_P(\boldsymbol{g}(\theta_s), \Lambda_s)$ & \\
        & $\boldsymbol{B}_i \sim \mathcal{N}(0,\Gamma_i)$ & $g_{p}(\theta_s) = a_s + \log \int_{\mathcal{E}_p} E^{-b_t + c_t \log E} {e^{-N_{\rm H} \sigma_p}}dE$ & 
        \multirow{2}{*}{$\sigma_i^2\sim$
        \text{IG}$(3,2\nu_i^2)$}
        \\
        & $\Gamma_i = \text{diag}(\boldsymbol{\tau}_i)R_i\text{diag}(\boldsymbol{\tau}_i)$ & $a_s \sim \text{non-informative prior}$ & \\
        & & $b_t \sim U(1.5, 3),\ c_t \sim U(-0.3, 0.5)$ & \\
        \bottomrule
    \end{tabular}
    \label{tab:model_priors}
\end{table*}

In the `base' variant, the hyperparameters $a$, $b$, and $c$ are scalar and do not differ by source. In the `source' variant, we consider separate values for the scaling hyperparameter $a_s$ indexed by source $s$ to capture source-specific scaling (brightness). The `target' variant additionally considers target-specific values for $b_t$ and $c_t$. Note that each of the 45 sources is associated with only one of the four AGN targets; this means specifying the source $s$ uniquely determines its original target $t$. In other words, we can express the target index as a function of the source index. %
\begin{definition}[\label{def:model_3_ext}Extension of Model 3]
    We extend Model 3 with sub-cases by varying the prior hyperparameters of the log-flux, $\boldsymbol{G}_s$. {Specifically, we consider} different structural choices for the prior covariance matrix $\Lambda_s$, namely Model (3A) $\Lambda_s=\psi^2I_P$, (3B) $\Lambda_s=\psi^2M_P$, (3C) $\Lambda_s=\psi_s^2I_P$, (3D) $\Lambda_s=\psi_s^2M_P$, where $I_P$ is the identity matrix of size $P$, $M_P$ is the first-corder auto-regressive correlation matrix of size $P$, and $\psi^2,\psi_s^2\sim\text{Inverse-Gamma}(6,0.5)$.
    Furthermore, the hyperparameter ${\theta}$ of the mean $\boldsymbol{g}_p({\theta})$ can vary depending on the source $s$ and the target $t$. The parameter ${\theta}_s$ has three variants: (Base) ${\theta}_s=(a,b,c)$, (Source) ${\theta}_s=(a_s,b,c)$, and (Source-target) ${\theta}_s=(a_s,b_t,c_t)$.
\end{definition}
The extensions of Model 3 are constructed based on prior knowledge of the dependency structures of astrophysical quantities. All three models and their corresponding variants are summarized in Table \ref{tab:model_priors}. {We refer to the variants of Models~3A, 3B, and 3C by appending the abbreviations `b', `s', and `st' for the Base, Source, and Source-targets variants, respectively, e.g., Model~3C-st.}
To identify the most suitable model, we fit all models to our photon count data and evaluate their relative performance in Section~\ref{sec:waic}.

%%%%%%%%%%%%%%%%%%%%%%%%%%%%%%%%%%%%%%%%%%%%%%%%%%%%%%%%%%%%%%%%%%%%%%%%%%%%%%%%%%%%%%%%%%%%%%%%%%%%%%%%%%%%%%%%%%%%%%%%%%%%%%%%%%%%%%%%%%%%%%%%%%%%%%

\section{Model Evaluation}\label{sec:model_evaluation}
In this section, we justify our choice of model structures, followed by a brief introduction to model comparison using information criteria. We begin by establishing theoretical foundations for our modeling framework.

%%%%%%%%%%%%%%%%%%%%%%%%%%%%%%%%%%%%%%%%%%%%%%%%%%

\subsection{Posterior Propriety And Parameter Identifiability}\label{sec:identifiability}

The model described in \eqref{eqn:log_model} is structurally not identifiable since for any given set of parameter values, one may introduce passband-specific constants $c_p$ such that 
$\tilde{B}_{ip}=B_{ip}-c_{p}$ and $\tilde{G}_{sp}=G_{sp}+c_{p}$ for each passband $p$. Under this transformation, the sum $B_{ip} + G_{sp}$ remains unchanged. This invariance induces non-identifiability at the likelihood level, commonly referred to as likelihood non-identifiability. We address this by applying constraints with instrument and source-specific priors. As a result, for posterior inference, it is necessary to verify the posterior identifiability of the model parameters under the proposed prior structures.

To establish the posterior identifiability, first we need to show that the posterior is proper, which is essential for meaningful inference and  valid posterior sampling, e.g., in the context of Hamiltonian Monte Carlo (HMC) methods. In Models 1 and 2, the log-flux is assigned a non-informative (and improper) prior, which may lead to an improper posterior distribution for the model parameters. To verify the propriety of the full posterior distribution $p(\boldsymbol{B}, \boldsymbol{G}, \boldsymbol{\sigma}^2 \mid \boldsymbol{y})$, we need to check that
\begin{equation}\label{eqn:posterior_check}
    \int p(\boldsymbol{B},\boldsymbol{G},\boldsymbol{\sigma}^2\mid \boldsymbol{y})\,d\boldsymbol{B}\,d\boldsymbol{G}\,d\boldsymbol{\sigma}^2 < \infty,
\end{equation}
where $\boldsymbol{\sigma}^2=(\sigma_1^2,\ldots,\sigma_I^2)$, and $\boldsymbol{B},\boldsymbol{G}$ are defined as concatenated vectors $\boldsymbol{B}=(\boldsymbol{B}_1,\ldots,\boldsymbol{B}_I)$ and $\boldsymbol{G}=(\boldsymbol{G}_1,\ldots,\boldsymbol{G}_S)$. Proposition \ref{prop:proper} proves posterior propriety for Models 1 and 2 with the mild condition that there is at least one observation for each source-passband combination.
\begin{proposition}\label{prop:proper}
The posterior distribution $p(\boldsymbol{B},\boldsymbol{G},\boldsymbol{\sigma}^2\mid \boldsymbol{y})$ is proper, provided that there is at least one observation for each combination of source-passband in the likelihood.
\end{proposition}
In Model 3, we have assigned conjugate and/or informative priors at all levels in the hierarchical model, ensuring that the posterior distribution of all model parameters is proper by construction. 
\begin{table*}
    \centering
    \caption{Comparison of WAIC across all model variants. Lower WAIC value indicates better predictive performance. All versions of Model 3 have better performance than Model 1. Among the Model 3 variants, Models 3A and 3C generally perform better than Models 3B and 3D. Model 3C-st achieves the best overall WAIC performance and is highlighted in bold.}
    \begin{tabular}{c|c|c|c|c|c|c|c}
    \midrule
        \multirow{4}{*}{Models}
        &  & \multicolumn{6}{c}{Model variants} \\
        \cmidrule{3-8}
        & Covariance structure
        & \multicolumn{2}{c|}{Base}
        & \multicolumn{2}{c|}{Source}
        & \multicolumn{2}{c}{Target} \\
        \cmidrule{3-8}
        &
        & WAIC$_1$ & WAIC$_2$
        & WAIC$_1$ & WAIC$_2$
        & WAIC$_1$ & WAIC$_2$ \\
        \midrule

        \multicolumn{2}{c|}{1}
        & $-10040.26$ & $-9793.21$
        &  &
        &  &  \\
        \cmidrule{1-4}

        \multicolumn{2}{c|}{2}
        & $-10047.40$ & $-9802.50$
        &  &
        &  &  \\
        \midrule

        3A & homogeneity
        & $-10047.79$ & $-9803.74$
        & $-10048.03$ & $-9803.66$
        & $-10046.85$ & $-9803.36$ \\
        \cmidrule{1-8}

        3B & homogeneity $+$ correlation
        & $-10046.44$ & $-9802.84$
        & $-10046.24$ & $-9802.25$
        & $-10046.47$ & $-9802.72$ \\
        \cmidrule{1-8}

        3C & heterogeneity
        & $-10047.92$ & $-9803.72$
        & $-10047.80$ & $-9803.06$
        & $\boldsymbol{-10048.50}$ & $\boldsymbol{-9803.90}$ \\
        \cmidrule{1-8}

        3D & heterogeneity $+$ correlation
        & $-10046.10$ & $-9802.29$
        & $-10046.46$ & $-9802.51$
        & $-10045.73$ & $-9801.69$ \\
        \bottomrule
    \end{tabular}
    \label{tab:waic}
\end{table*}

We turn to identifiability of the model parameters, which ensures uniqueness of posterior inference. For posterior identifiability, we need to show that the joint posterior $p(\boldsymbol{B}, \boldsymbol{G}, \boldsymbol{\sigma}^2 \mid \boldsymbol{y})$ admits a unique global maximizer.
We proceed in two steps. First, we prove posterior identifiability of $\boldsymbol{\sigma}^2$ by differentiating the joint posterior likelihood with respect to $\boldsymbol{\sigma}^2$. Proposition 3 establishes that given data $\boldsymbol{y}$, we can uniquely identify the posterior maximum for $\boldsymbol{\sigma}^2$.
\begin{proposition}
    \label{prop:identifiability_sigma}
Given a fixed set of observations $\boldsymbol{y}$, the the joint log-posterior
\begin{align*}
    \ell(\boldsymbol{\sigma}^2\mid \boldsymbol{B},\boldsymbol{G},\boldsymbol{y})
&=\log p(\boldsymbol{y}\mid \boldsymbol{B},\boldsymbol{G},\boldsymbol{\sigma}^2)+\log p(\boldsymbol{B},\boldsymbol{G}\mid \boldsymbol{\sigma}^2)\\
&\hspace{2.8cm}+\log p(\boldsymbol{\sigma}^2)+\text{constant}
\end{align*}
is maximized at a fixed value of $\boldsymbol{\sigma}$ that satisfies the normalizing equation \eqref{eqn:identifiability_maxima}.
\end{proposition}
This directly establishes the posterior identifiability for $\boldsymbol{\sigma}^2$, but identifiability of $\boldsymbol{B}$ and $\boldsymbol{G}$, requires Proposition \ref{prop:identifiability}.
\begin{proposition}\label{prop:identifiability}
Given $\boldsymbol{\sigma}^2$, the conditional posterior distribution $p({\boldsymbol{B}}, {\boldsymbol{G}}\, | \,\boldsymbol{\sigma}^2, \boldsymbol{y})$ of $(\boldsymbol{B},\boldsymbol{G})$ is a multivariate Gaussian distribution $\mathcal{N}\big(\boldsymbol{\mu}_{BG} ( \boldsymbol{\sigma}^2,\boldsymbol{y}), \Sigma_{BG}( \boldsymbol{\sigma}^2,\boldsymbol{y})\big)$, where $\Sigma_{BG}( \boldsymbol{\sigma}^2,\boldsymbol{y})$ is a positive definite matrix.
\end{proposition}
Combining Propositions \ref{prop:identifiability_sigma} and \ref{prop:identifiability}, the joint posterior admits a unique global maximizer, establishing posterior uniqueness.

%%%%%%%%%%%%%%%%%%%%%%%%%%%%%%%%%%%%%%%%%%%%%%%%%%%%%%%%

\subsection{Model Selection With Information Criterion}\label{sec:waic}
To compare the posterior performance of the model variants, we use the \textit{Watanabe-Akaike Information Criterion (WAIC)}, introduced in \citet{watanabe2010asymptotic}, as our metric for model evaluation. In the hierarchical modeling framework, several parameters are informed by a limited subset of the data. Hence, the effective number of parameters in the model is an important consideration when selecting an appropriate model. WAIC evaluates a model's posterior predictive accuracy while adjusting for its complexity (i.e., the effective number of parameters), as discussed in \citet{gelman2014understanding}. This measure is computed using \textit{log pointwise predictive density} (lppd) and the \textit{effective number of parameters} ($p_{WAIC}$). Suppose a model with parameter $\boldsymbol{\eta}$ is fitted to data $\boldsymbol{y}$ with $n$ observations, and we obtain a posterior sample $\{\boldsymbol{\eta}^{(j)}: j=1,\ldots,J\}$ of the parameter $\eta$ from its posterior distribution $p(\boldsymbol{\eta}\mid\boldsymbol{y})$. WAIC is defined in one of two ways,
\[\mathrm{WAIC}_1=-2 \cdot \mathrm{lppd}+2 \cdot p^{(1)}_{\mathrm{WAIC}} \ \hbox{ or } \  
\mathrm{WAIC}_2=-2 \cdot \mathrm{lppd}+2 \cdot p^{(2)}_{\mathrm{WAIC}}
\]
where 
\[\text { lppd }=\sum_{i,s,p} \log \left(\frac{1}{J} \sum_{j=1}^J p\left(y_{isp} \mid \boldsymbol{\eta}^{(j)}\right)\right),\]
and $p_{\text{WAIC}}$ is computed in two ways: $p^{(1)}_{\text {WAIC }}$ is
\begin{align*}
    &2 \sum_{i,s,p}\left(\log \bigg(\frac{1}{J} \sum_{j=1}^J p\left(y_{isp} \mid \boldsymbol{\eta}^{(j)}\right)\bigg) -\frac{1}{J} \sum_{j=1}^J \log p\left(y_{isp} \mid \boldsymbol{\eta}^{(j)}\right)\right)
\end{align*}
and (let $\text{Var}_{j=1}^J$ denote sample variance):
\[p^{(2)}_{\text {WAIC }}=\sum_{i,s,p} \text{Var}_{j=1}^J\left(\log p\left(y_{isp} \mid\boldsymbol{\eta}^{(j)}\right)\right).\]

The two definitions of $p_{\text{WAIC}}$ yield slightly different effective numbers of parameters. A lower WAIC value indicates a better-suited model for capturing data variability.

In our setting, the posterior predictive distribution of ${y}_{isp}$ given $B_{ip}$, $G_{sp}$, $\sigma_{i}$ is $\mathcal{N}(y_{isp};- {\sigma}_i^2/2 + {B}_{ip} + {G}_{sp},\,{\sigma}_i^2)$. Using a posterior Monte Carlo sample of the model parameters, $\boldsymbol{\eta}^{(j)}=(\hat{\boldsymbol{B}},\hat{\boldsymbol{G}},\hat{\boldsymbol{\sigma}}^2)^{(j)}$, $j=1,\ldots,J$; we define the posterior predictive density as \[p(y_{isp}\mid\boldsymbol{\eta}^{(j)}) =\frac{1}{\hat{\sigma}_i^{(j)}}\phi\left(\frac{y_{isp}+(\hat{\sigma}^{(j)}_i)^2/2-\hat{B}^{(j)}_{ip}-\hat{G}^{(j)}_{sp}}{\hat{\sigma}^{(j)}_i}\right)\]
and
\[\text { lppd }=\sum_{i,s,p} \log \left(\frac{1}{J} \sum_{j=1}^J p\left(y_{isp} \mid\boldsymbol{\eta}^{(j)}\right)\right),\]
where $\phi$ is the standard Gaussian probability density function. Using the posterior sample of the parameters ${B}_{ip}$, ${G}_{sp}$, and ${\sigma}_i^2$ obtained via the Hamiltonian Monte Carlo (HMC) from \texttt{STAN}, we evaluate the posterior predictive distribution for all available data, $y_{isp}$. The summations in this section run across all available instrument–source–passband combinations. These likelihood values are then used to compute WAIC for candidate models and their variants.

%%%%%%%%%%%%%%%%%%%%%%%%%%%%%%%%%%%%%%%%%%%%%%%%%%%%%%%%%%%%%%%%%%%%%%%%%%%%%%%%%%%%%%%%%%%%%%%%%%%%%%%%%%%%%%%%%%%%%%%%%%%%%%%%%%%%%%%%%%%%%%%%%%%%%

\section{Data Analysis And Results}\label{sec:results}

In this section, we evaluate and compare the performance of the proposed models using WAIC and posterior predictive behavior. We then present detailed visualizations of the posterior estimates for relevant parameters to facilitate a comprehensive comparison between model variants. Supporting plots and figures are presented to illustrate empirical findings. We conclude with a discussion of the feasibility of identifying time-dependent patterns in estimates of relative error of effective area using current data.

%%%%%%%%%%%%%%%%%%%%%%%%%%%%%%%%%%%%%%%%%%%%%%%%%%

\subsection{Model Comparison}\label{sec:model_comparison}

We fit the models and compare their posterior performance. For example, when we compare the posterior predictive mean square error, Model 3C-st exhibits one of the lowest residual mean squared errors (RMSE) among the competing models.  However, a low RMSE value alone does not imply a more useful model, as RMSE does not account for model complexity or the effective number of parameters. Thus, we select a model using the WAIC values in Table~\ref{tab:waic}. The results indicate that Model~1 exhibits higher WAIC values than the other models, suggesting that modeling the passband-wise correlation for the effective area parameter $B_{ip}$ enables more accurate capture of the data variation. Additionally, all variants in Model 3 achieve lower WAIC values than Model 2, suggesting that an informative Gaussian prior with a specified hierarchical structure on the log-flux parameter $G_{sp}$ (like in Definition~\ref{def:model_2}) further improves model performance.

Next, we compared the variants of Model 3 (described in Definition \ref{def:model_3_ext}) with the aim of capturing a range of plausible model structures that are hypothesized by experts. Table~\ref{tab:waic} shows that incorporating prior correlation across energy passbands for the log-flux parameter $G_{sp}$ does not appear to improve model performance. Specifically, Models 3A and 3C, which do not include passband-wise correlation in the log-flux, consistently achieve lower WAIC values than Models 3B and 3D, which do not impose such correlation. It is important to note that passband-wise correlation in the effective area parameter $B_{ip}$ is included in all variants of Model 3 and appears to improve model performance. In contrast, imposing a first-order autoregressive structure on the log-flux prior does not yield similar benefits.

Comparing Models 3A and 3C, both have similar WAIC values in the base and source-indexed variants. But Model 3C performs substantially better in the target-indexed variants. This implies that allowing source-specific variance $\psi_s^2$ in $\boldsymbol{G}_s$ (Definition \ref{def:model_3_ext}) improves the performance of the model. Overall, the source and target-indexed variant of Model 3C (Model 3C-st) yields the lowest WAIC under both definitions of WAIC.

\begin{figure}
  \centering
  \includegraphics[scale=0.45]{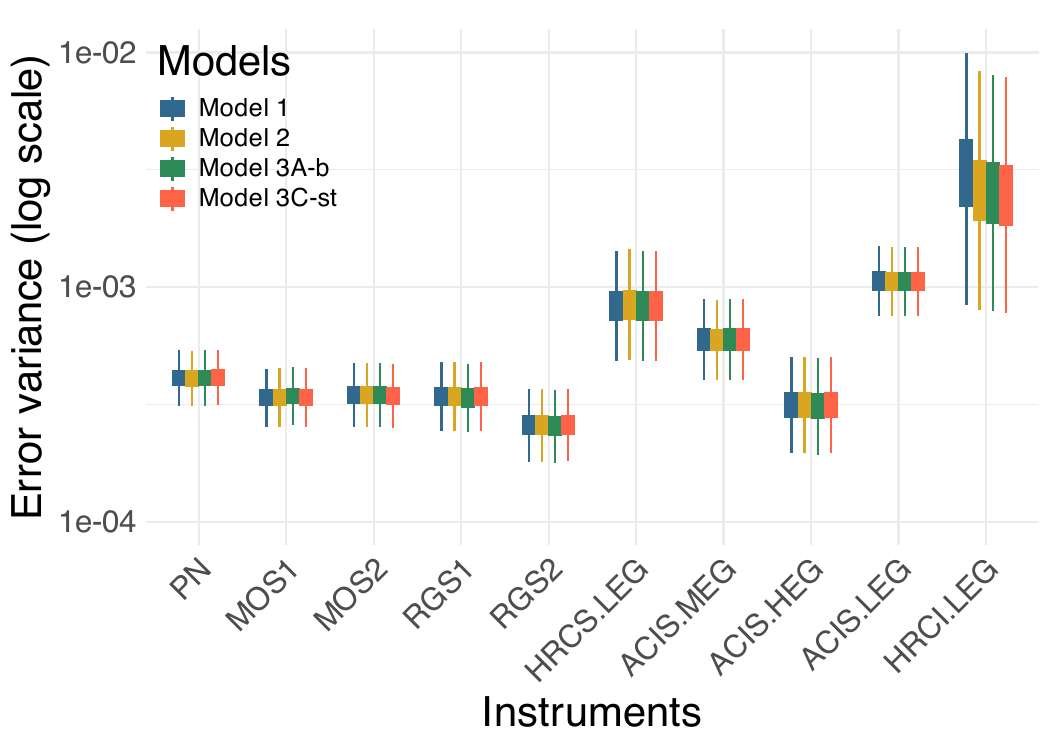}
  \caption{Boxplots of posterior sample of the model error variance, $\sigma_i^2$, across ten instruments for four models.  
  The vertical axis is plotted on a logarithmic scale.}
  \label{fig:sigma_plot}
\end{figure}

Consistent with this comparison, Model 3C-st also produces lower posterior estimates of the variance parameter $\sigma_i^2$. In particular, for the \texttt{HRCI.LEG} instrument, Model 1 has the highest variance estimate, while Model 3C-st has the lowest posterior estimates. Figure~\ref{fig:sigma_plot} presents boxplots of $\sigma_i^2$ for Models 1 and 2 and two variants for Model 3, Model 3A-base, which is the simplest variant of Model 3, and Model 3C-st, which is the best performing model. Based on the performance of all Model 3 variants, we consider Model 3C-st  to be our preferred choice for calibration concordance.

\begin{figure*}
\centering
\begin{minipage}{0.48\textwidth}
\centering
\fig{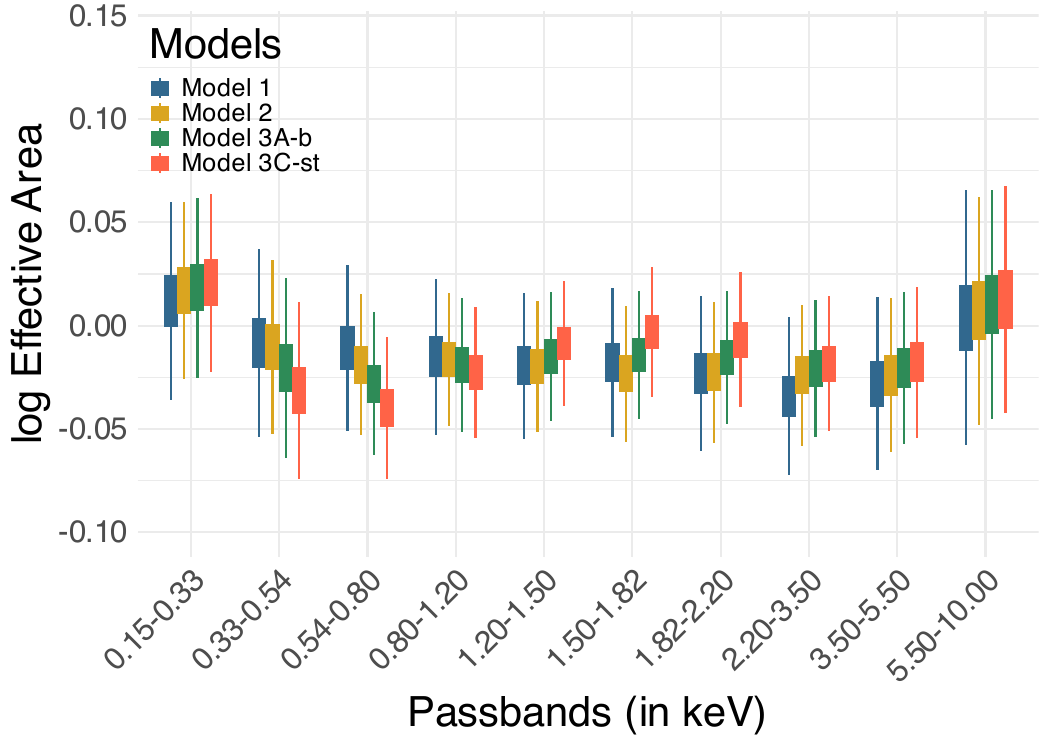}{\linewidth}{(a) Instrument \texttt{PN}}
\end{minipage}
\begin{minipage}{0.48\textwidth}
\centering
\fig{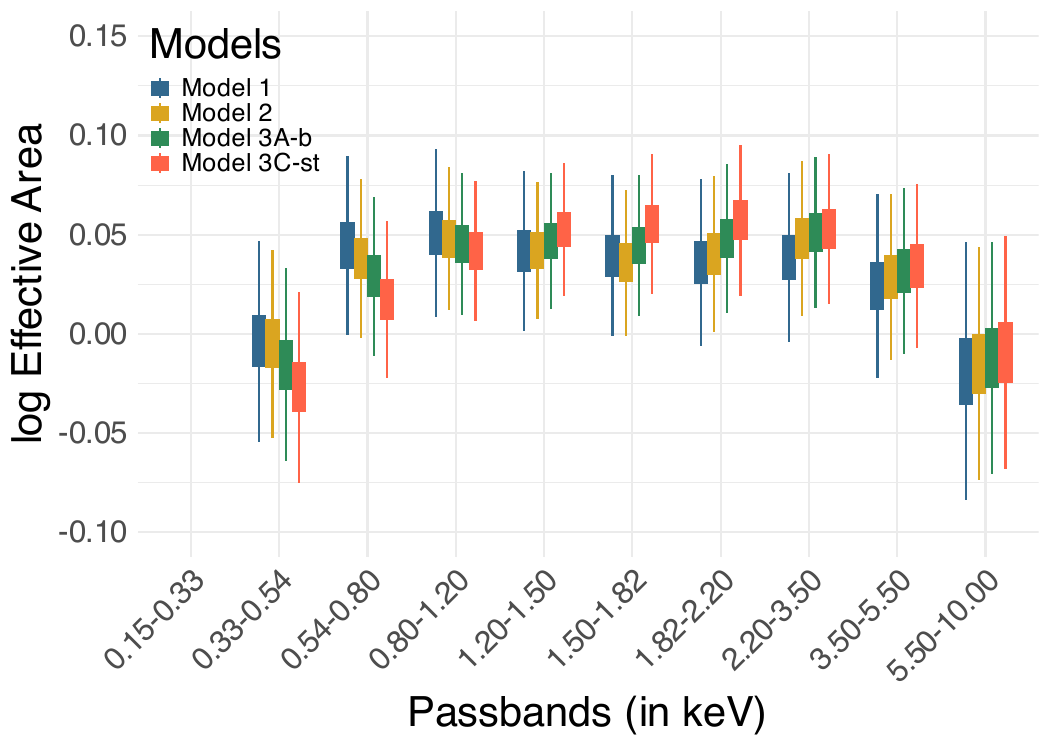}{\linewidth}{(b) Instrument \texttt{ACIS.LEG}}
\end{minipage}
\caption{Boxplots of simulated posterior distributions of the relative error of effective area ${B}_{ip}$ across ten passbands for instruments (a) \texttt{PN} and (b) \texttt{ACIS.LEG}, under four models. The horizontal axis denotes ten energy passbands. For each passband, boxplots of posterior distributions are plotted for four Models 1, 2, 3A-b and 3C-st. For the \texttt{ACIS.LEG} instrument, the posterior distributions of $B_{ip}$ are missing in the lowest passband (0.15-0.33 keV), since there are no observation available for this particular instrument-passband combination. The parameter estimates  are within the tolerance boundary of $\pm$0.20.}
\label{fig:instrument_plot}
\end{figure*}

\begin{figure*}
\gridline{
  \fig{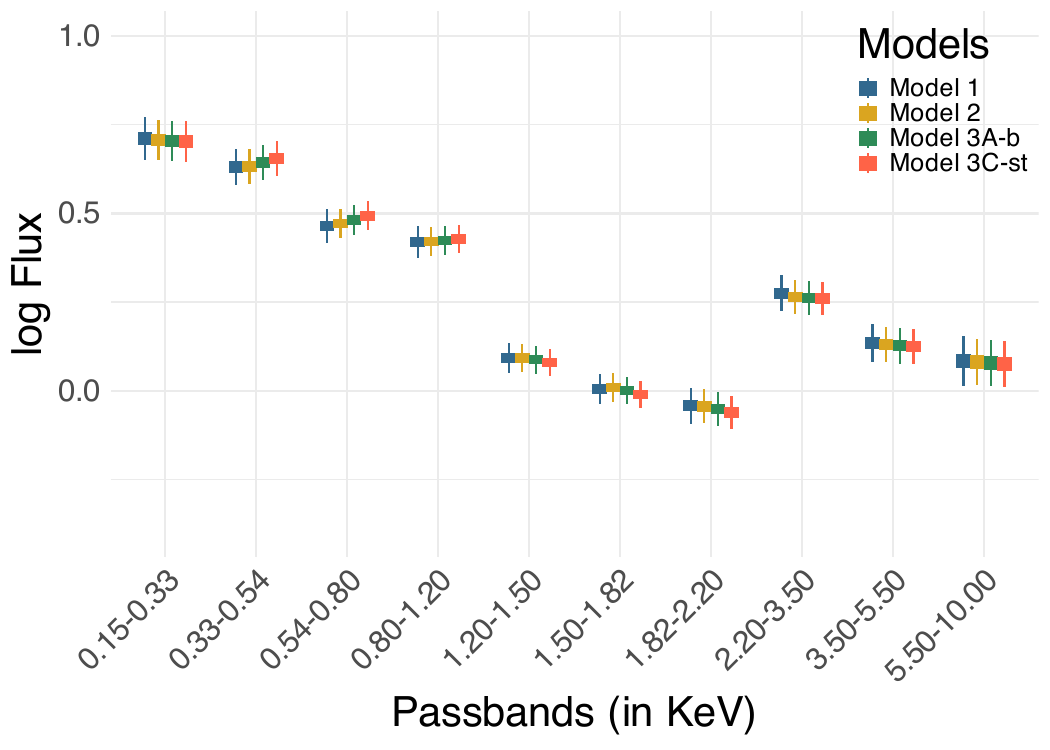}{0.45\linewidth}{(a) Source \texttt{1703}}
  \fig{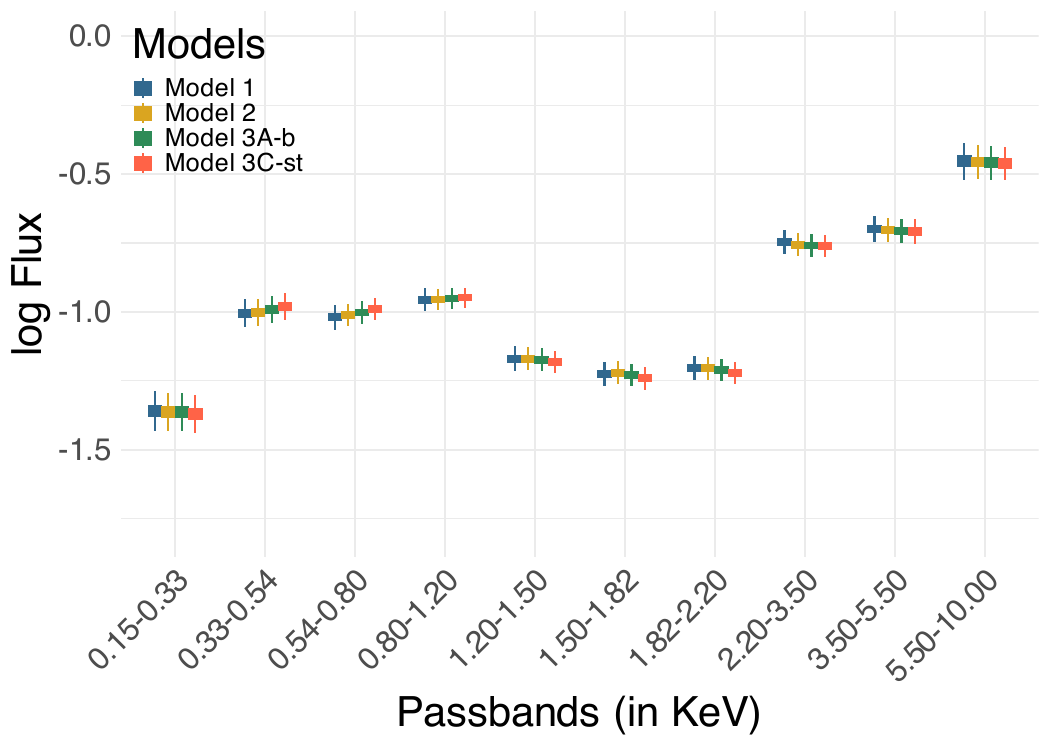}{0.45\linewidth}{(b) Source \texttt{4924A}}
}
\caption{Boxplots of simulated posterior distributions of the log-flux, $G_{sp}$, across ten passbands for two sources (a) \texttt{1703} and (b) \texttt{4924A}, and four models.}
\label{fig:source_plot}
\end{figure*}

The box plots of the logarithmic effective area parameter $B_{ip}$ for two instruments (\texttt{PN} and \texttt{ACIS.LEG}) are presented in Figure \ref{fig:instrument_plot}. Plots for the other instruments are given in Appendix~\ref{appx:plots}. As mentioned in Section~\ref{sec:log_transformed_model}, a tolerance threshold of $\pm0.20$ in the effective area error $B_{ip}$ corresponds to a 20\% measurement error for the effective area on the original scale. The majority of the estimates fall within the specified tolerance range, and the boxplots contain zero values, validating the effectiveness of our models' estimation capability.

The posterior distributions of the log-flux parameter $G_{sp}$ appear as boxplots in Figure~\ref{fig:source_plot}, for two randomly selected sources (\texttt{1703} and \texttt{4924A}). The magnitudes of the log-flux estimates are source-specific. In some sources, the flux estimates increase with energy passband, whereas in others they decrease. This pattern originates from the shape of the underlying spectral distribution of the photon energy for each source.

%%%%%%%%%%%%%%%%%%%%%%%%%%%%%%%%%%%%%%%%%%%%%%%%%%

\subsection{Inference of Model Parameters}\label{sec:parameter_plots}

\begin{figure}
  \centering
\includegraphics[scale=0.45]{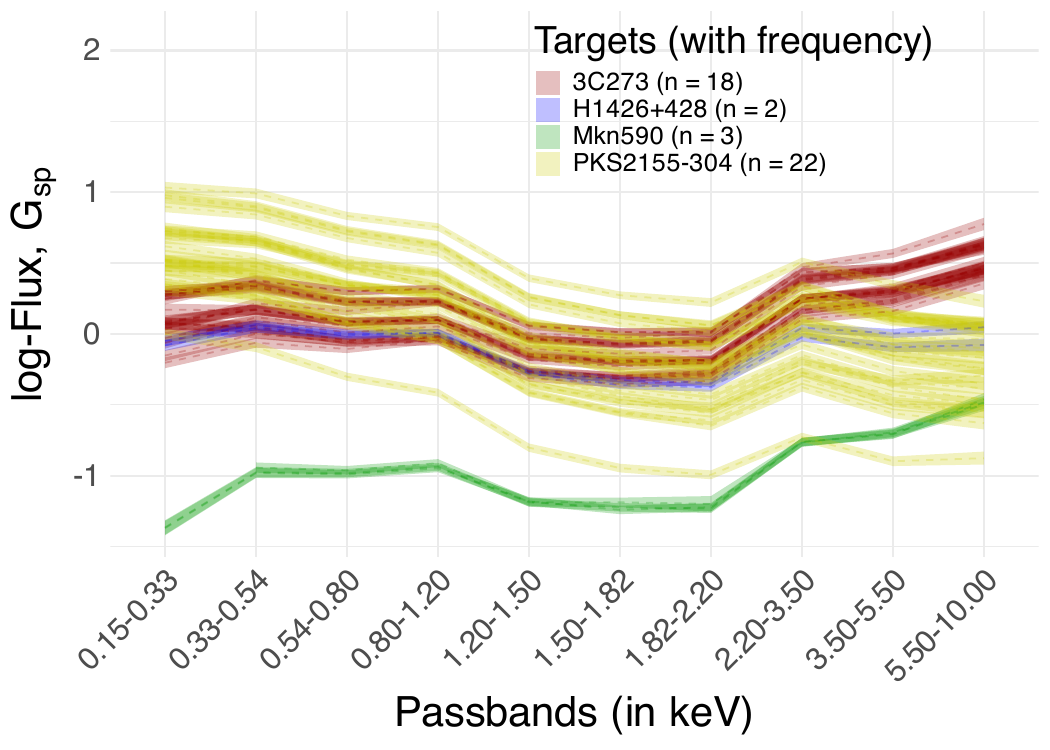}
  \caption{Posterior summaries of the log-flux parameter $G_{sp}$ across passbands ordered by increasing energy level. For each source, the posterior mean and the lower and upper bounds of the 95\% credible interval are connected across passbands. The lines are colored according to the AGN target from which the source originates. Sources associated with the same target exhibit similar, approximately parallel log-flux profiles across energy passbands, with increasing or decreasing trends depending on the target. The legend contains, in parentheses, the number of sources associated with each target.}
  \label{fig:target_plot}
\end{figure}

Besides having the lowest WAIC, Model 3C-st includes separate source and target-specific hyperparameters in the prior mean of log-flux $G_{sp}$. This allows posterior inference on the target-specific spectral parameters. In Figure \ref{fig:target_plot}, we plot 95\% credible interval of the posterior log-flux ${G}_{sp}$ across increasing energy passbands for all sources. When we color them by target, we notice target-specific pattern in the posterior estimate of log-flux parameter $G_{sp}$. The log-flux interval boundaries of sources from the same target are largely parallel to each other. This underscores the importance of considering target specific parameter and source-specific scaling parameter in the log-flux prior structure. In this section, we discuss how Model 3C-st facilitates the estimation of source and target-specific parameters without sacrificing posterior predictive capacity.

\begin{figure*}
\centering
\gridline{\fig{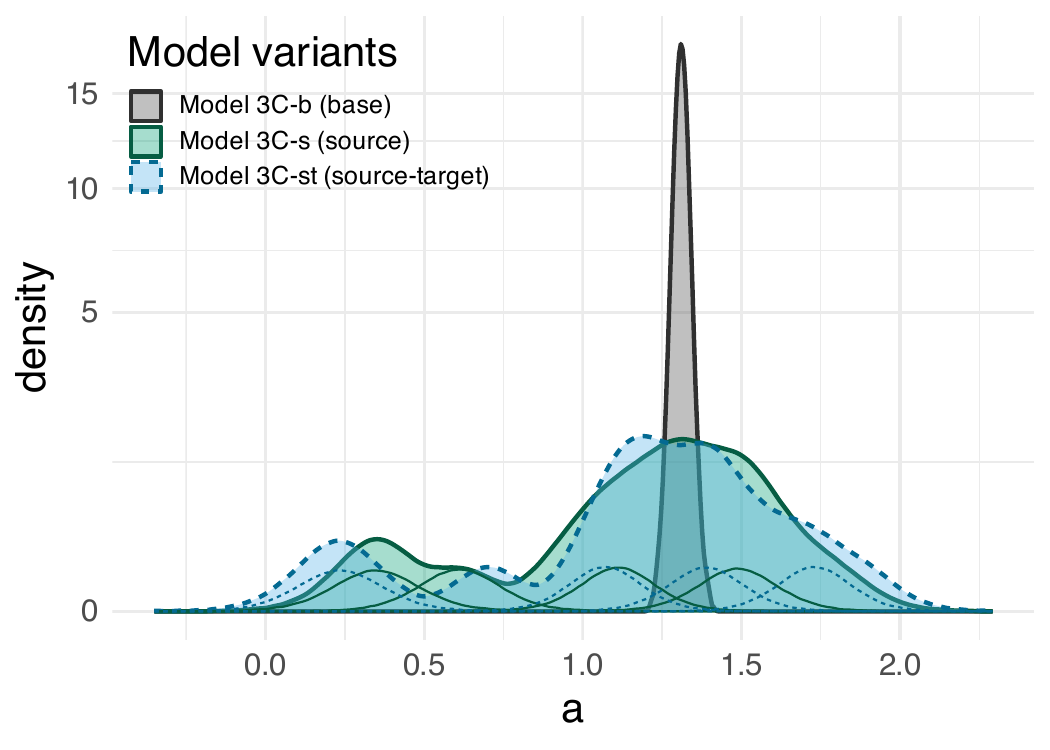}{0.45\textwidth}{(a)}}
\gridline{
  \fig{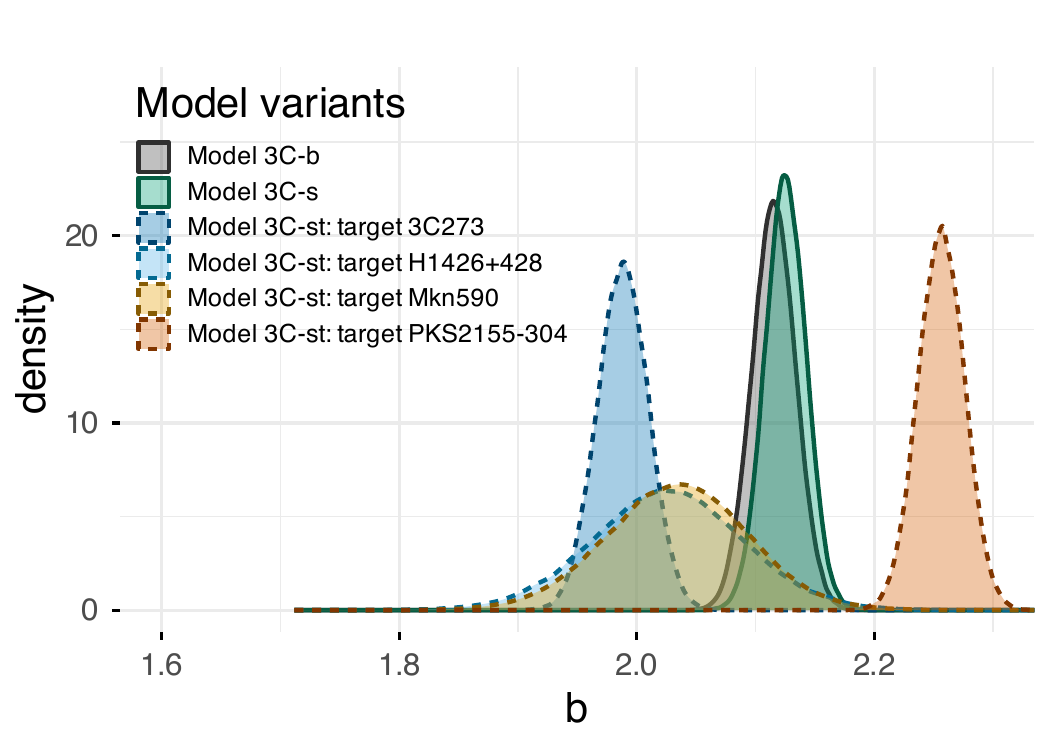}{0.45\textwidth}{(b)}\hfill
  \fig{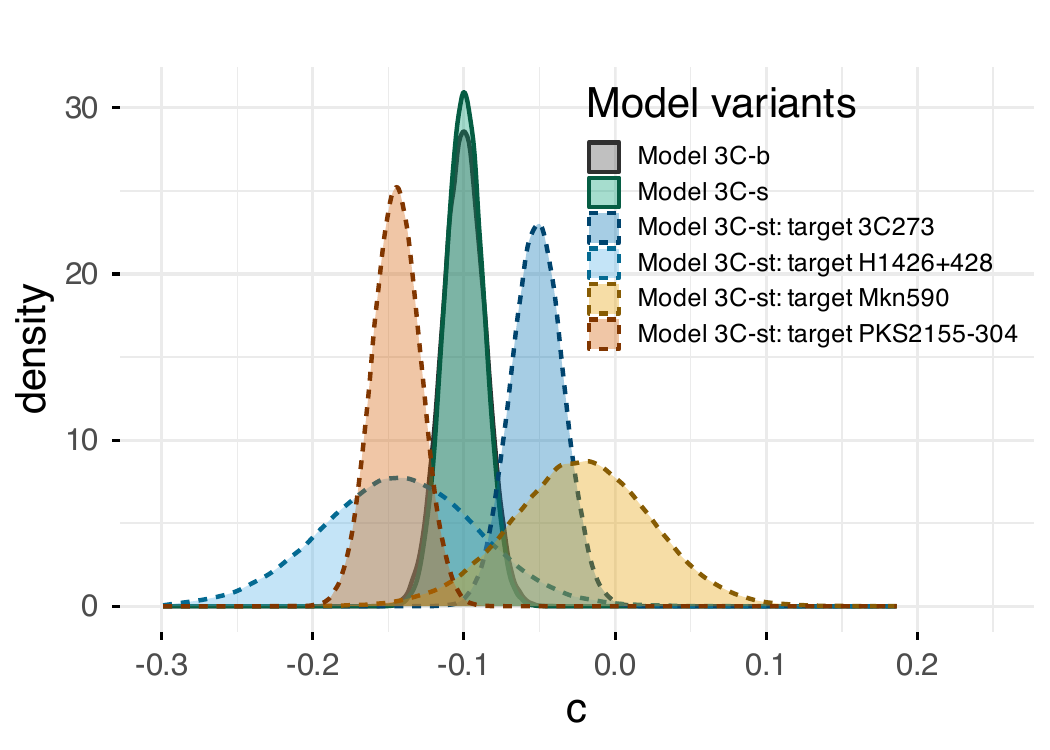}{0.45\textwidth}{(c)}
}
\caption{Posterior density estimates of the hyperparameters in the prior for log-flux ${G}_{sp}$ under the three variants of Model 3C. Panel (a) shows the posterior densities of the hyperparameter $a$. In Models 3C-s and 3C-st, $a$ varies by source $s$, denoted by $a_s$; whereas in Model 3C-b $a$ is common across all sources. For Models~3C-s and 3C-st, the bold lines represent the mixture density across the source-specific components. The thin lines show selected individual posterior densities of $a_s$ for several sources. Panels (b) and (c) show the posterior densities of the hyperparameters $b$ and $c$, respectively. In Model 3C-st these parameters vary by target $t$, denoted by $b_t$ and $c_t$; while in Models 3C-b and 3C-s they are common across targets. Allowing $a_s$ to vary by source substantially increases the spread of the posterior relative to the shared model, reflecting greater source-level variability. This behavior is largely unaffected by whether $b$ and $c$ are estimated separately for each target. In contrast, the posterior distributions of $b$ and $c$ are nearly identical in Models 3C-b and 3C-s, but differ markedly when $b_t$ and $c_t$ are estimated separately for each target in Model 3C-st.}
\label{fig:abc_plot}
\end{figure*}

Figure~\ref{fig:abc_plot} presents plots of the posterior densities of $a$, $b$ and $c$, corresponding to the three variants of Model 3C.  In the base Model 3C-b, $a$ is a shared scalar paarmeter with a unimodal posterior distribution. But in Model 3C-s and 3C-st, $a$ is allowed to vary by source, yielding source-specific parameters $a_s$ for each of the 45 sources. For these two models, the mixture density of parameters $a_s$ is plotted aggregating over all sources, which is significantly different from that in Model 3C-b. Allowing $a_s$ to vary across sources substantially widens the span of the (aggregate) posterior distribution relative to the shared index model, reflecting greater source-level variability.

In Model 3C-st, the hyperparameters $b$ and $c$ also vary by target $t$, denoted by $b_t$ and $c_t$; whereas in Models 3C-b and 3C-s, they are treated as index-free scalars across targets.  The posterior estimates of the parameters $b$ and $c$ change significantly when the model includes target-indexed parameters. It is important to note that the posterior distribution of $a$ remains largely unaffected by whether $b$ and $c$ are indexed by the target; similarly, the indexation of $a$ does not affect the posterior distributions for $b$ and $c$. Such behavior of posterior distributions indicates that our model can effectively disentangle the information governing these parameters. To assess the robustness of the model fit, all models have been refitted to partial data with a subset of instruments and sources.

\begin{figure*}
\gridline{
  \fig{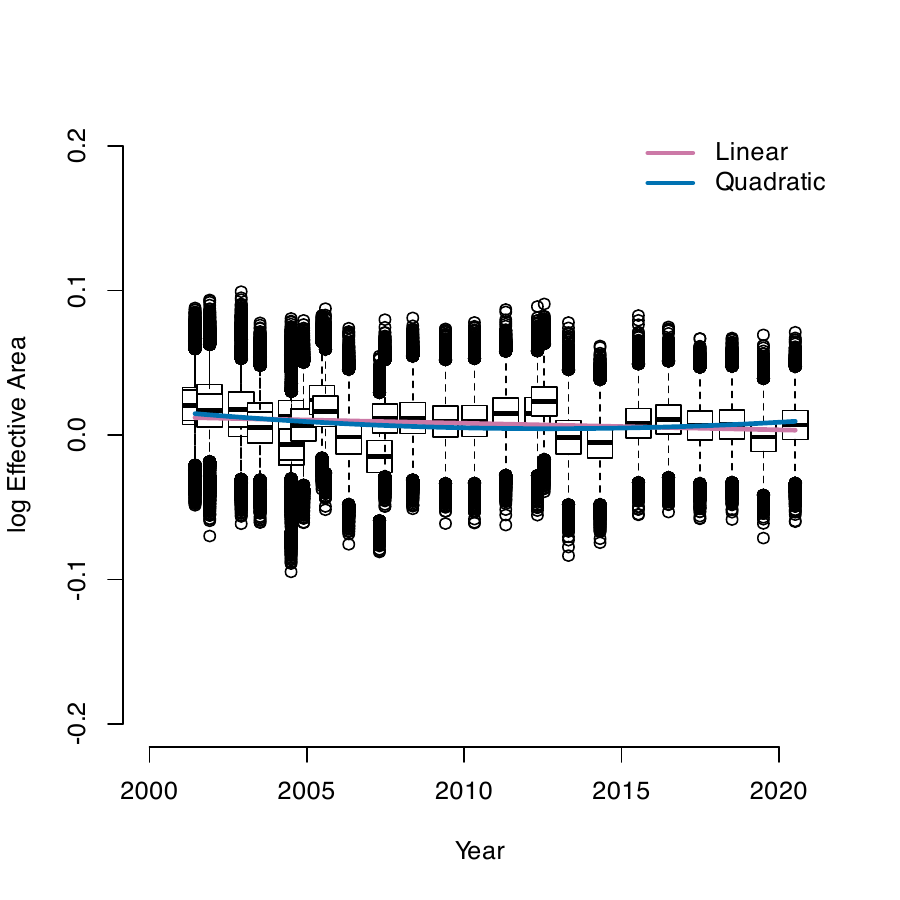}{0.45\textwidth}{(a) Instrument \texttt{MOS1}, Passband 1.82--2.20 keV}
  \fig{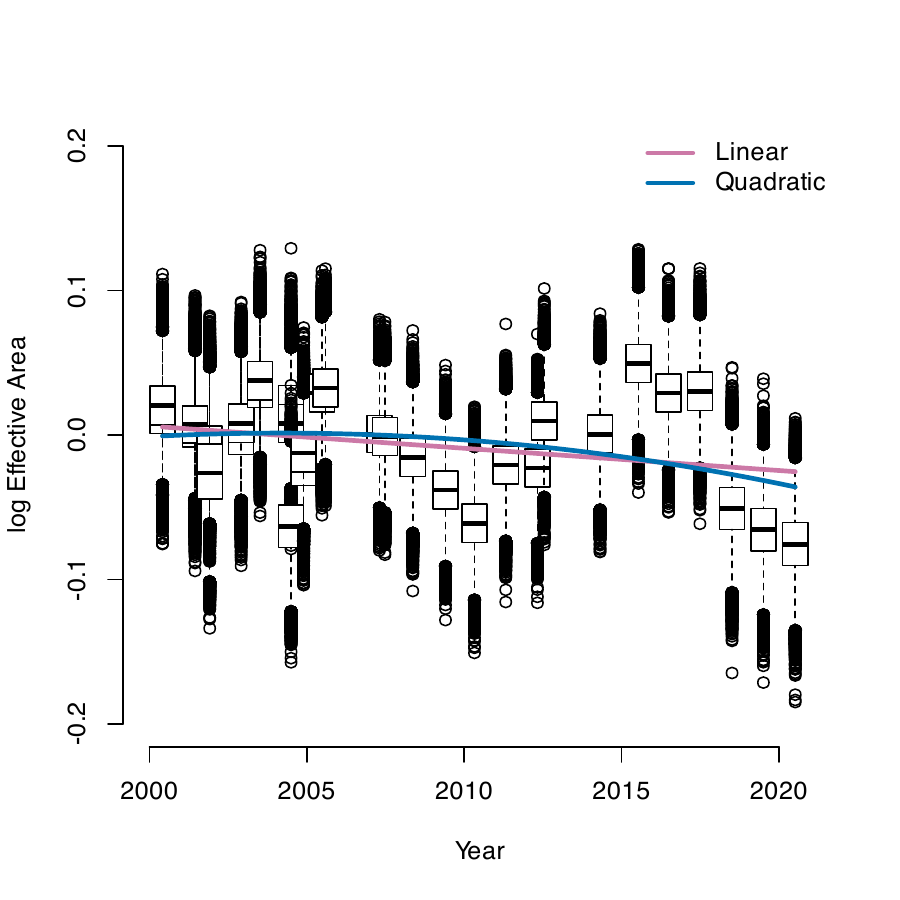}{0.45\textwidth}{(b) Instrument \texttt{MOS2}, Passband 0.15--0.33 keV}
}
\caption{Boxplots of posterior samples of relative error estimates of effective areas plotted against the corresponding time of photon measurements (in decimal years since 2000). Panel (a) shows results for instrument \texttt{MOS1} and passband 1.82--2.20~keV, while panel (b) corresponds to instrument \texttt{MOS2} and passband 0.15--0.33~keV. Fitted linear and quadratic regression curves of the effective area relative error against photon collection time are also superimposed.}
\label{fig:timeplot}
\end{figure*}

%%%%%%%%%%%%%%%%%%%%%%%%%%%%%%%%%%%%%%%%%%%%%%%%%%

\subsection{Evolution Of Error in Effective Area Over Time}\label{sec:time}

In this observational study we used photon flux data collected from multiple instruments measured at different time. Each astronomical source is associated with a timestamp, defined as the `elapsed time' between the year $2000$ and the time of the photon measurement. We can use this information to explore the temporal trend, increasing or decreasing, of an instrument's effective area errors. The recorded time is specific to the source being measured, hence we can denote it by $t(s)$, a function of the particular source $s$. Then for every source $s$, the error in effective area can be estimated as a function of time $t$ by
$$\hat{B}_{ip}(t):=\hat{B}_{ip}(t(s)) =y_{isp}+\frac{1}{2}\hat{\sigma}_i^2-\hat{G}_{sp}.$$
This formulation enables us to estimate a time-dependent version of the effective area deviation. In Figure~\ref{fig:timeplot}, we present boxplots of the posterior distribution of the effective area error estimates for two specific instrument-passband combinations, plotted against the timing of photon measurements. We also fit a linear and a quadratic polynomial regression curve for the simulated effective area $\hat{B}_{ip}(t)$ against time $t$ for each of these plots. In Table~\ref{tab:regression_coef}, we present the coefficient estimates of a linear fit between the logarithm of the effective area against time. We observe no strong or systematic temporal trend in the effective area error for many of the passbands, while for others, we can see a clear linear trend. A more detailed investigation with a larger data set is necessary to rigorously assess any potential time dependence with statistical significance.

\begin{table*}
\centering
\caption{Linear regression slope estimates between logarithmic effective area and observation year for all instrument and passband combinations. Entries report the estimated slope and three times its regression standard error in units of $10^{-3}\,\mathrm{dex\,yr^{-1}}$. A dash indicates that fewer than two distinct time points were available.}
\label{tab:regression_coef}
\begin{tabular}{lccccc}
\toprule
\multirow{2}{*}{Instrument} & \multicolumn{5}{c}{Passbands (in keV)}\\
\cmidrule{2-6}
 & $0.15$--$0.33$ & $0.33$--$0.54$ & $0.54$--$0.80$ & $0.80$--$1.20$ & $1.20$--$1.50$ \\
\midrule
\texttt{PN} & 4.25 $\pm$ 0.009 & 1.56 $\pm$ 0.006 & 1.29 $\pm$ 0.005 & 0.94 $\pm$ 0.004 & 0.29 $\pm$ 0.004 \\
\texttt{MOS1} & -3.66 $\pm$ 0.012 & 1.65 $\pm$ 0.006 & 0.54 $\pm$ 0.005 & -0.11 $\pm$ 0.004 & -0.58 $\pm$ 0.004 \\
\texttt{MOS2} & -1.49 $\pm$ 0.010 & 1.60 $\pm$ 0.006 & 0.01 $\pm$ 0.005 & 0.17 $\pm$ 0.004 & 0.23 $\pm$ 0.004 \\
\texttt{RGS1} & \textemdash & -1.61 $\pm$ 0.006 & -1.15 $\pm$ 0.004 & -0.59 $\pm$ 0.006 & -0.62 $\pm$ 0.005 \\
\texttt{RGS2} & \textemdash & -1.51 $\pm$ 0.005 & -0.94 $\pm$ 0.005 & -0.20 $\pm$ 0.004 & 0.26 $\pm$ 0.004 \\
\texttt{HRCS.LEG} & 6.69 $\pm$ 0.042 & 4.52 $\pm$ 0.030 & 3.40 $\pm$ 0.026 & 7.10 $\pm$ 0.024 & 4.68 $\pm$ 0.028 \\
\texttt{ACIS.MEG} & \textemdash & 7.00 $\pm$ 0.035 & 3.34 $\pm$ 0.011 & 0.36 $\pm$ 0.006 & 0.92 $\pm$ 0.006 \\
\texttt{ACIS.HEG} & \textemdash & \textemdash & \textemdash & -0.10 $\pm$ 0.006 & -0.03 $\pm$ 0.006 \\
\texttt{ACIS.LEG} & \textemdash & -11.49 $\pm$ 0.017 & -0.30 $\pm$ 0.011 & -0.03 $\pm$ 0.008 & 0.79 $\pm$ 0.008 \\
\texttt{HRCI.LEG} & \textemdash & \textemdash & \textemdash & \textemdash & \textemdash \\
\midrule
\midrule
\multirow{2}{*}{Instrument} & \multicolumn{5}{c}{Passbands (in keV)}\\
\cmidrule{2-6}
 & $1.50$--$1.82$ & $1.82$--$2.20$ & $2.20$--$3.50$ & $3.50$--$5.50$ & $5.50$--$10.00$ \\
\midrule
\texttt{PN} & 1.25 $\pm$ 0.005 & 0.70 $\pm$ 0.005 & 0.28 $\pm$ 0.005 & 0.21 $\pm$ 0.005 & -0.17 $\pm$ 0.008 \\
\texttt{MOS1} & -1.12 $\pm$ 0.005 & -0.51 $\pm$ 0.005 & -0.69 $\pm$ 0.005 & -0.05 $\pm$ 0.005 & 0.69 $\pm$ 0.008 \\
\texttt{MOS2} & -0.01 $\pm$ 0.005 & -0.36 $\pm$ 0.005 & 0.12 $\pm$ 0.005 & 0.25 $\pm$ 0.005 & -0.31 $\pm$ 0.008 \\
\texttt{RGS1} & -0.30 $\pm$ 0.006 & \textemdash & \textemdash & \textemdash & \textemdash \\
\texttt{RGS2} & -0.49 $\pm$ 0.005 & \textemdash & \textemdash & \textemdash & \textemdash \\
\texttt{HRCS.LEG} & 1.76 $\pm$ 0.028 & 8.05 $\pm$ 0.037 & 15.28 $\pm$ 0.037 & -3.06 $\pm$ 0.088 & \textemdash \\
\texttt{ACIS.MEG} & 1.07 $\pm$ 0.006 & 0.55 $\pm$ 0.007 & 0.50 $\pm$ 0.007 & 0.29 $\pm$ 0.009 & -1.11 $\pm$ 0.025 \\
\texttt{ACIS.HEG} & 0.82 $\pm$ 0.006 & -0.26 $\pm$ 0.007 & 0.04 $\pm$ 0.006 & -0.64 $\pm$ 0.007 & 0.37 $\pm$ 0.011 \\
\texttt{ACIS.LEG} & 1.34 $\pm$ 0.008 & 2.27 $\pm$ 0.009 & 1.24 $\pm$ 0.009 & -0.25 $\pm$ 0.013 & -0.99 $\pm$ 0.029 \\
\texttt{HRCI.LEG} & \textemdash & \textemdash & \textemdash & \textemdash & \textemdash \\
\bottomrule
\end{tabular}
\end{table*}

%%%%%%%%%%%%%%%%%%%%%%%%%%%%%%%%%%%%%%%%%%%%%%%%%%%%%%%%%%%%%%%%%%%%%%%%%%%%%%%%%%%%%%%%%%%%%%%%%%%%%%%%%%%%%%%%%%%%%%%%%%%%%%%%%%%%%%%%%%%%%%%%%%%%%%

\section{Summary}\label{sec:conclusion}
In this article, we establish an in-flight calibration standard using photon measurement data collected by multiple instruments observing various astronomical sources. We apply an additive log-normal model with a half-variance correction to the log-transformed photon counts. The models included energy passband levels in both instrument and source effects to capture passband-specific variations in the parameters, which were not accounted for in existing models in the literature. 

Empirical estimates with cross-passband correlations are used, along with hierarchical priors involving parametric spectrum model for source fluxes. We consider three versions of the prior structure to gradually introduce a parametric formulation of increasingly complex models. Relative model performance was evaluated using the Watanabe-Akaike Information Criterion (WAIC) to identify the best-performing model, adjusted for the effective number of parameters. We also examine the models' identifiability and assess the posterior propriety of the parameters. Finally, a comparative visualization of posterior estimates was presented through boxplots to understand the calibration requirement for a particular instrument.

The current parameter estimates obtained are specific to calibrating the ten instruments considered in our analysis. To calibrate a new instrument, photon measurements from the same set of sources using the new instrument are required. Then we can fit the model again with the augmented dataset that includes observations from the new instrument, and use the parameter estimates for further analyses.

Future work could explore richer prior structures. For instance, one might introduce a time-dependent polynomial term into the prior mean of the effective area error, $B_{ip}$. This would allow inference on the temporal trend of each instrument’s efficacy. However, such an extension would substantially increase model complexity at the risk of identifiability issues, potentially jeopardizing posterior convergence. Obtaining reliable posterior estimates for the polynomial coefficients under this framework would likely require a substantially larger dataset, which is beyond the scope of the present analysis.

\section*{Acknowledgements} 
Support for this work was provided in part by the National Aeronautics and Space Administration (NASA) through the Smithsonian Astrophysical Observatory (SAO) contract SV3-73016 to MIT for support of the Chandra X-Ray Center (CXC), which is operated by SAO for and on behalf of NASA under contract NAS8-03060. Support was also provided by NASA under contract NAS8-39073 to SAO. This work was conducted in collaboration with the CHASC International Astrostatistics Center and supported in part by the Engineering and Physical Sciences Research Council [grant number EP/W015080/1]; and the National Science Foundation under Grant Nos. [DMS-2113615, DMS-2113397, and DMS-2113605]. In addition, D.v.D. was supported in part by Marie Sk\l odowska-Curie RISE (H2020-MSCA-RISE-2019-873089) grants provided by the European Commission.
This is work is based on observations obtained with XMM-Newton, an ESA science mission with instruments and contributions directly funded by ESA Member States and NASA.
For the purpose of open access, the authors have applied a Creative Commons Attribution (CC BY) licence to any Author Accepted Manuscript version arising. Any opinions, findings, and conclusions or recommendations expressed in this material are those of the author(s) and do not necessarily reflect the views of the National Science Foundation. 

%%%%%%%%%%%%%%%%%%%%%%%%%%%%%%%%%%%%%%%%%%%%%%%%%%

\section*{Data Availability} 
The \textit{Chandra X-ray Observatory} data used in this article are available via the Chandra Data Collection (CDC) \href{https://doi.org/10.25574/cdc.693}{doi:10.25574/cdc.693}; the XMM-\textit{Newton} data is available via the
\href{https://nxsa.esac.esa.int/nxsa-web/}{XMM-Newton Science Archive}
or
\href{https://heasarc.gsfc.nasa.gov/W3Browse/}{HEASARC W3Browse}.

%%%%%%%%%%%%%%%%%%%%%%%%%%%%%%%%%%%%%%%%%%%%%%%%%%
%%%%%%%%%%%%%%%%%%%%%%%%%%%%%%%%%%%%%%%%%%%%%%%%%%

%\begin{thebibliography}{99}
%\end{thebibliography}

%%%%%%%%%%%%%%%%%%%%%%%%%%%%%%%%%%%%%%%%%%%%%%%%%%

%%%%%%%%%%%%%%%%% APPENDICES %%%%%%%%%%%%%%%%%%%%%

\appendix

%%%%%%%%%%%%%%%%%%%%%%%%%%%%%%%%%%%%%%%%%%%%%%%%%%

\section{Additional Results from Model Fitting}\label{appx:plots}

In this section, we provide more visualizations of our fitted models. The styles of the figures in this section are exactly the same as the figures in the main text and are self-explanatory. 

\begin{figure*}
\centering
\gridline{
  \fig{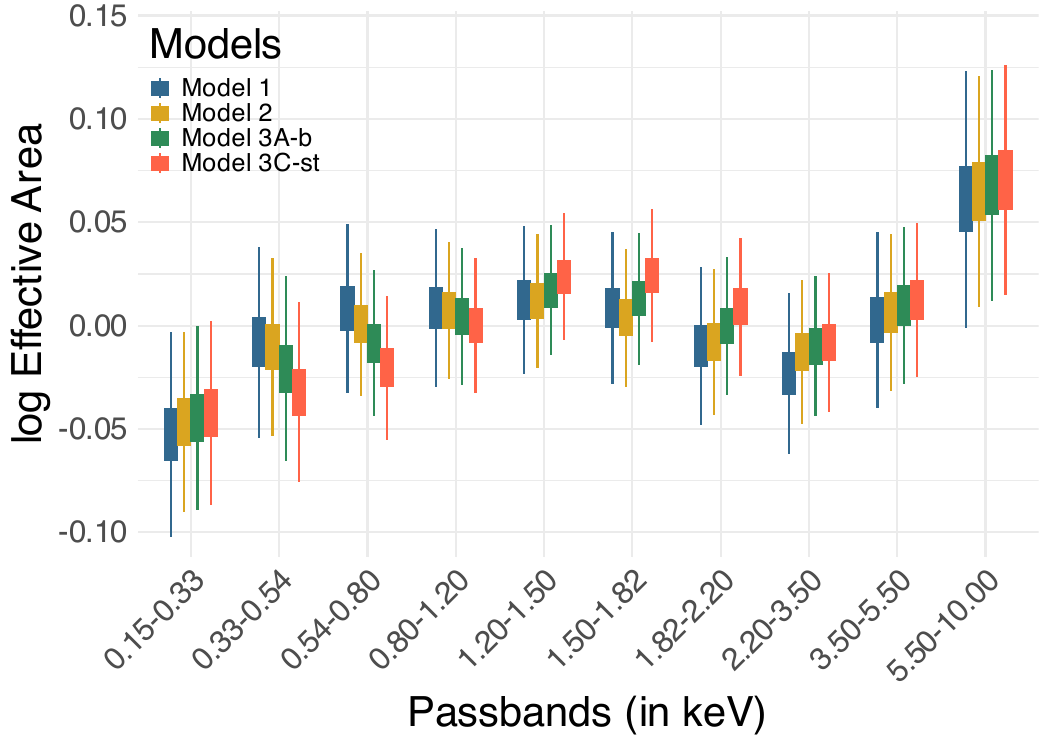}{0.33\textwidth}{(a) Instrument \texttt{MOS1}}\hfill
  \fig{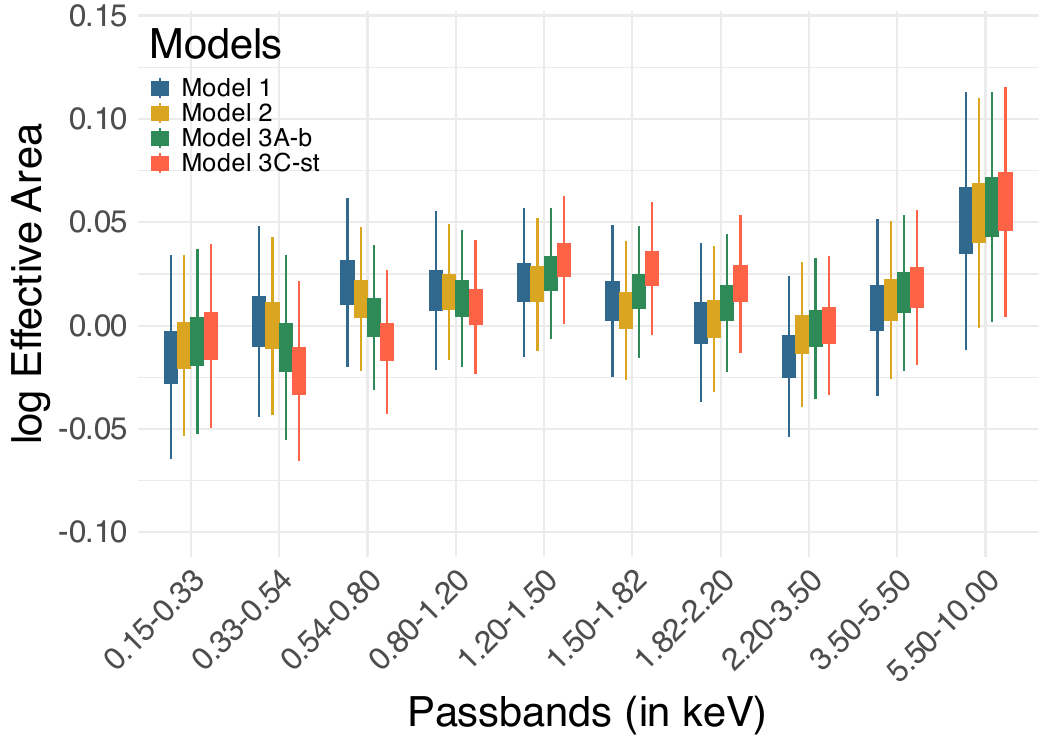}{0.33\textwidth}{(b) Instrument \texttt{MOS2}}\hfill
}
\gridline{
  \fig{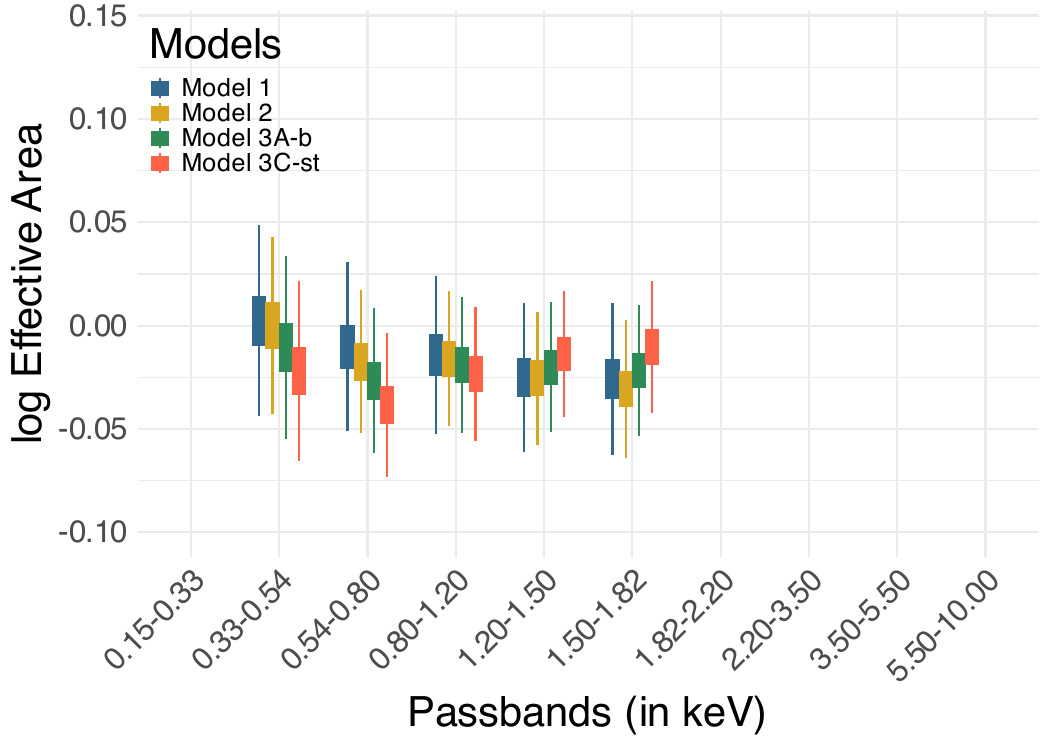}{0.33\textwidth}{(c) Instrument \texttt{RGS1}}\hfill
  \fig{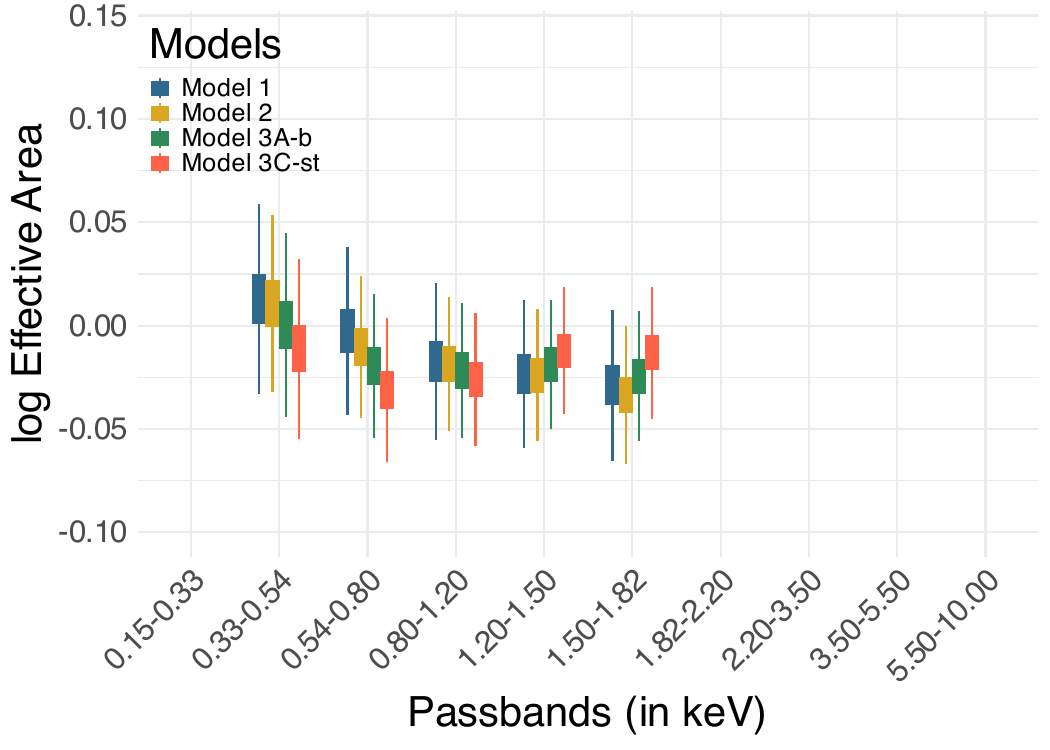}{0.33\textwidth}{(d) Instrument \texttt{RGS2}}\hfill
}
\gridline{
  \fig{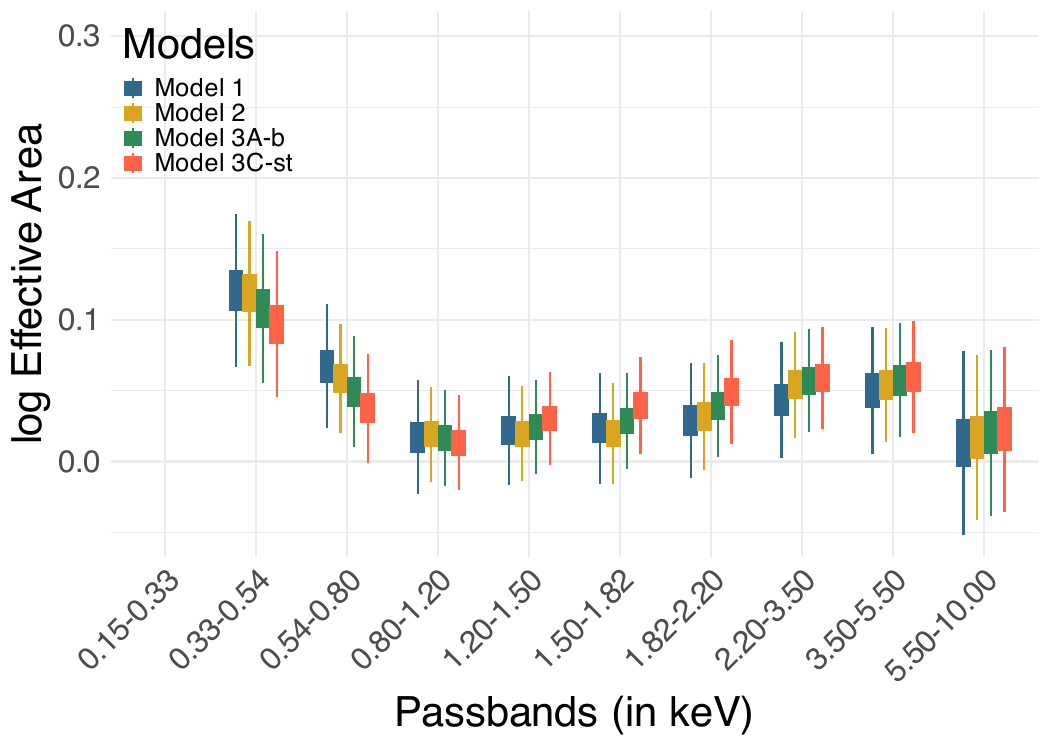}{0.33\textwidth}{(e) Instrument \texttt{ACIS.MEG}}
  \fig{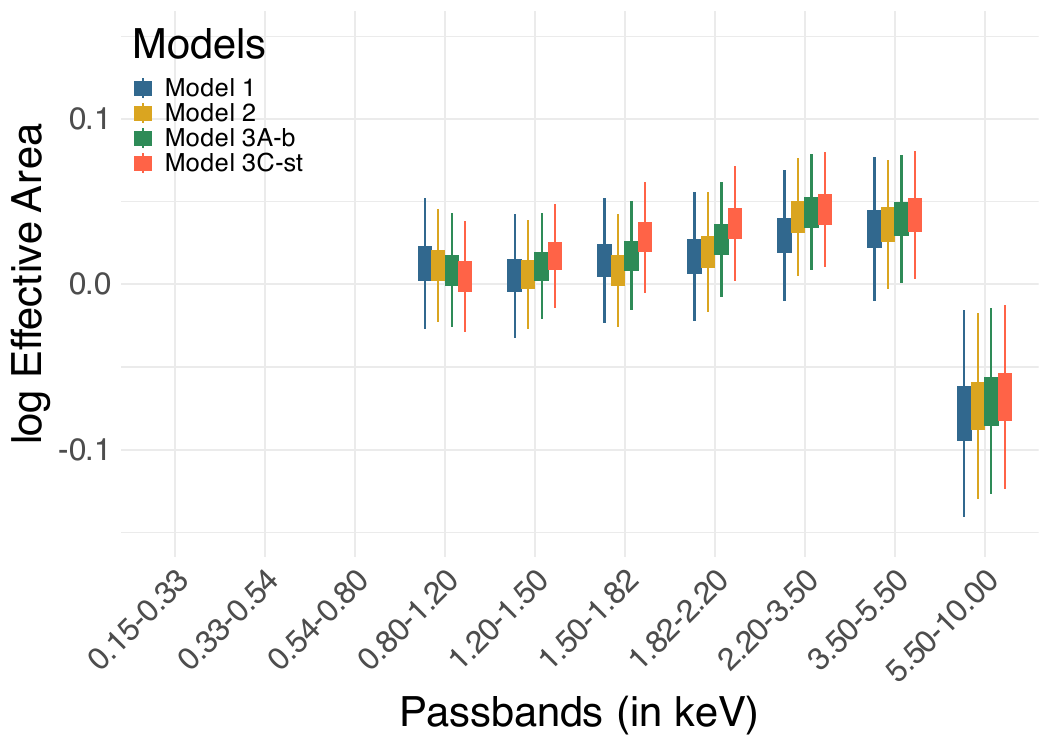}{0.33\textwidth}{(f) Instrument \texttt{ACIS.HEG}}
}
\gridline{
  \fig{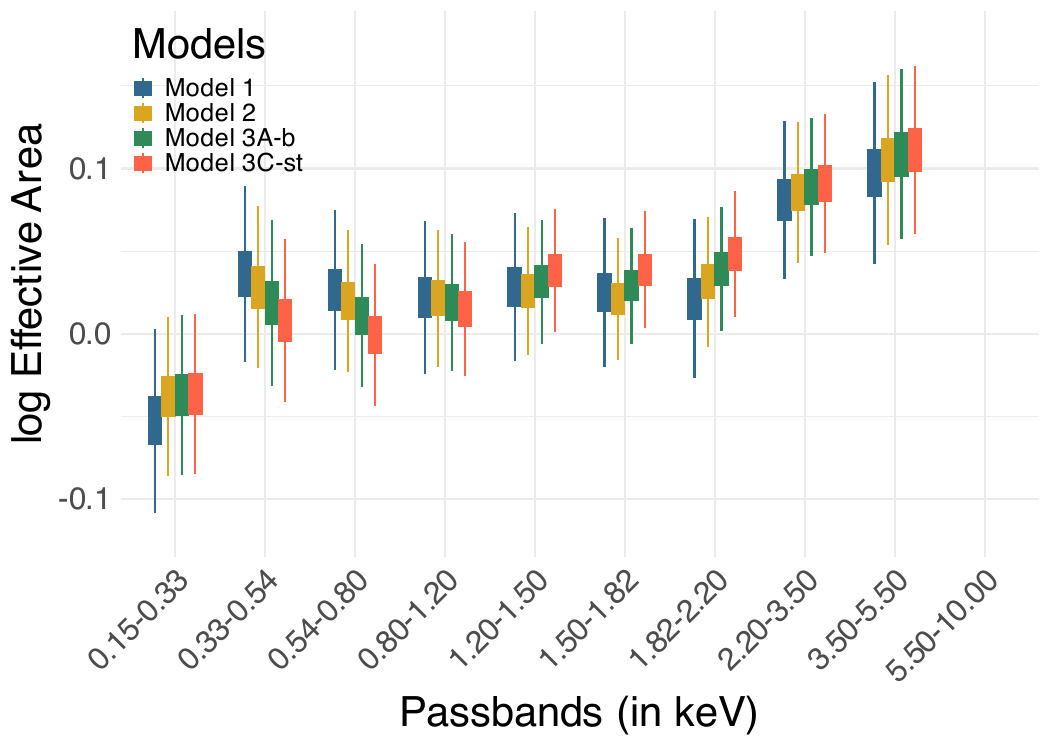}{0.33\textwidth}{(g) Instrument \texttt{HRCS.LEG}}
  \fig{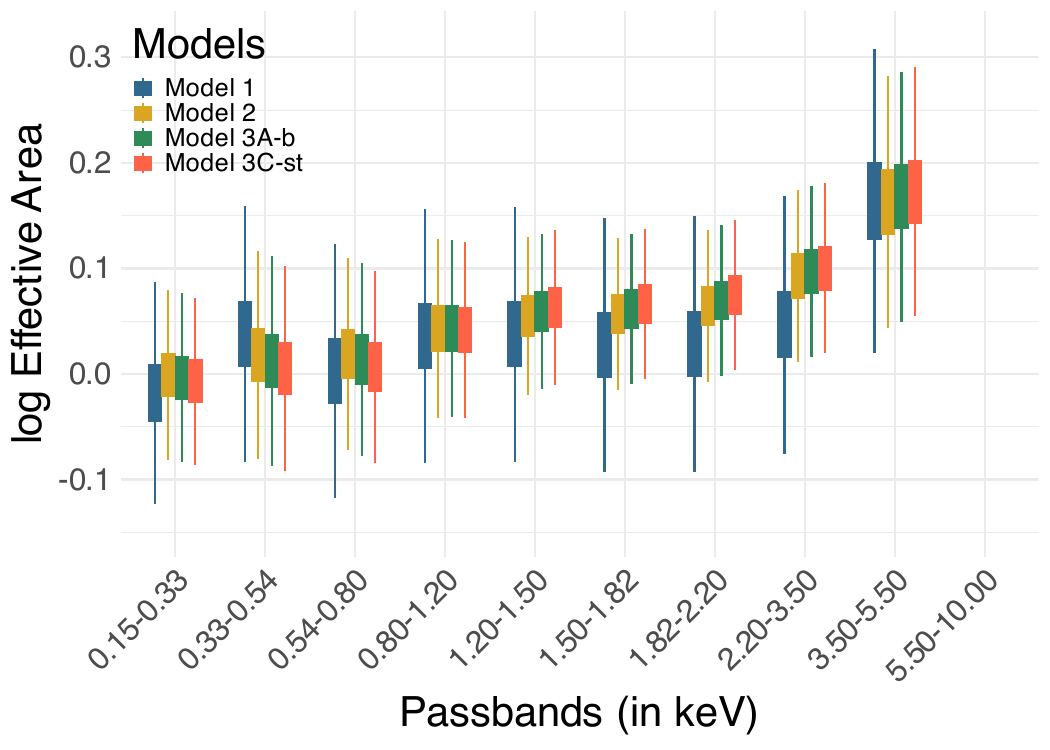}{0.33\textwidth}{(h) Instrument \texttt{HRCI.LEG}}
}
\caption{Boxplots of simulated posterior estimates of the effective area error parameter $\hat{B}_{ip}$ across ten passbands for eight instruments under four models 1, 2, 3A-b and 3C-st; plots for other instruments are showed in Figure~\ref{fig:instrument_plot}. For some instruments, certain passbands do not have any observation. Posterior estimates of $B_{ip}$ for those instrument-passband combinations are missing. The tolerance bounds of $\pm0.20$ are indicated in the corresponding panels, where visible within the plotting range. Only for the highest passband 5.50-10.0 keV in instrument \texttt{HRCI.LEG}, some of the estimate outliers exceed this boundary.}
\label{fig:appx_instrument_plot}
\end{figure*}

\begin{figure*}
\centering
\gridline{
  \fig{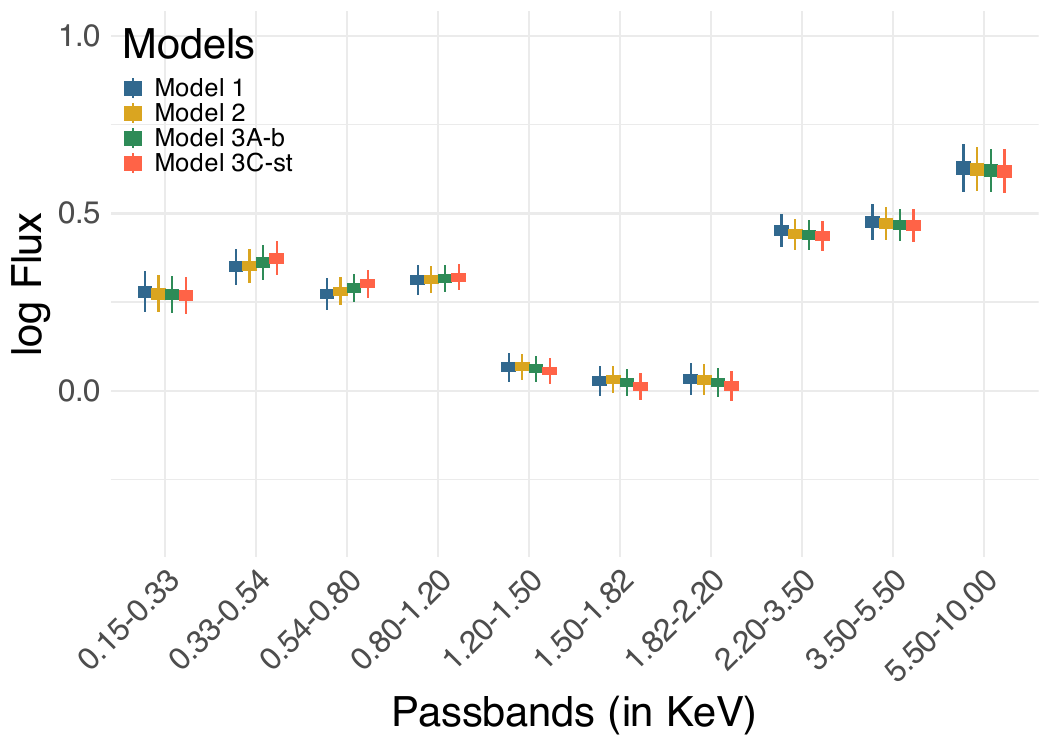}{0.33\textwidth}{(a) Source \texttt{4430}}\hfill
  \fig{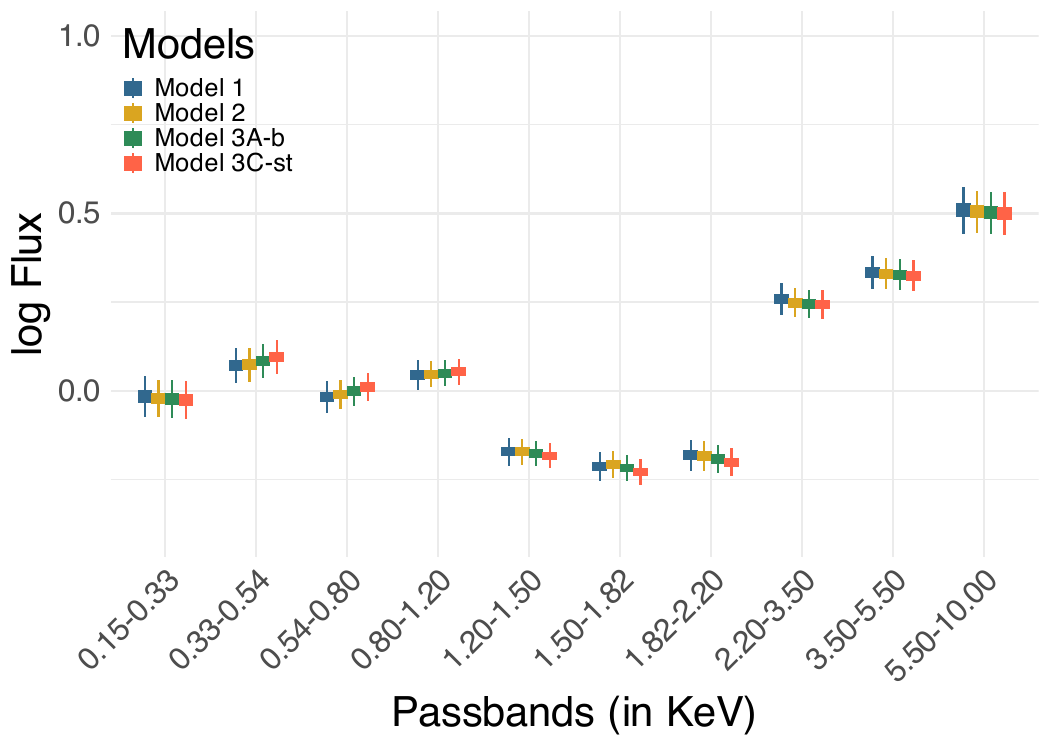}{0.33\textwidth}{(b) Source \texttt{14455}}\hfill
}
\gridline{
  \fig{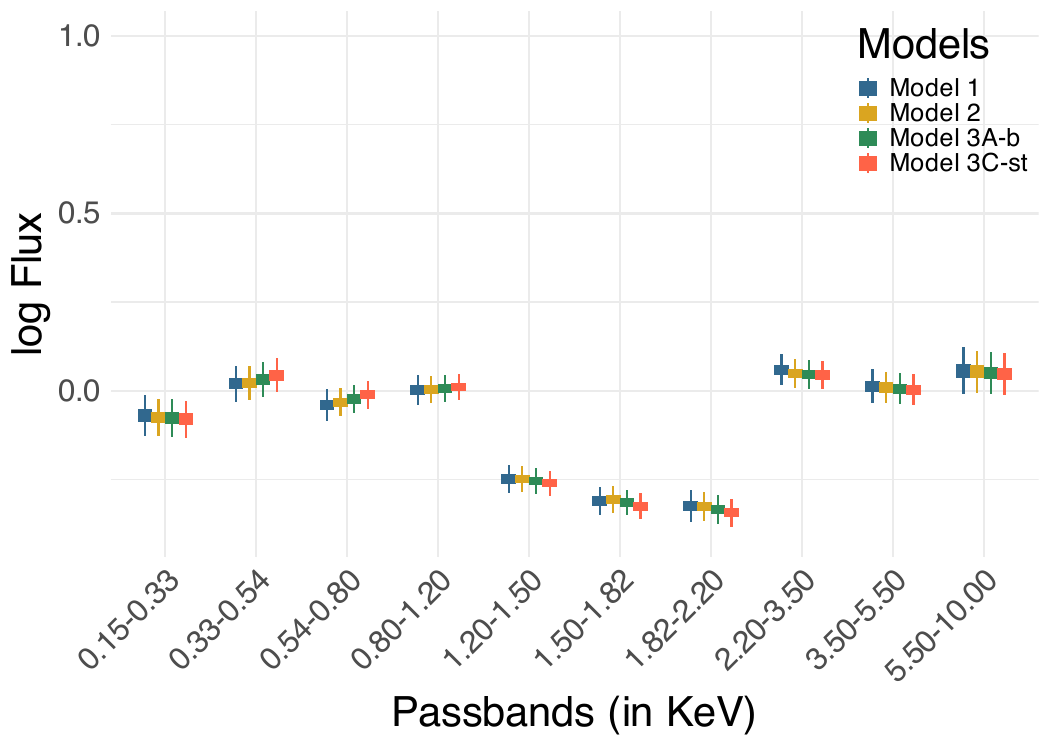}{0.33\textwidth}{(c) Source \texttt{6088}}\hfill
  \fig{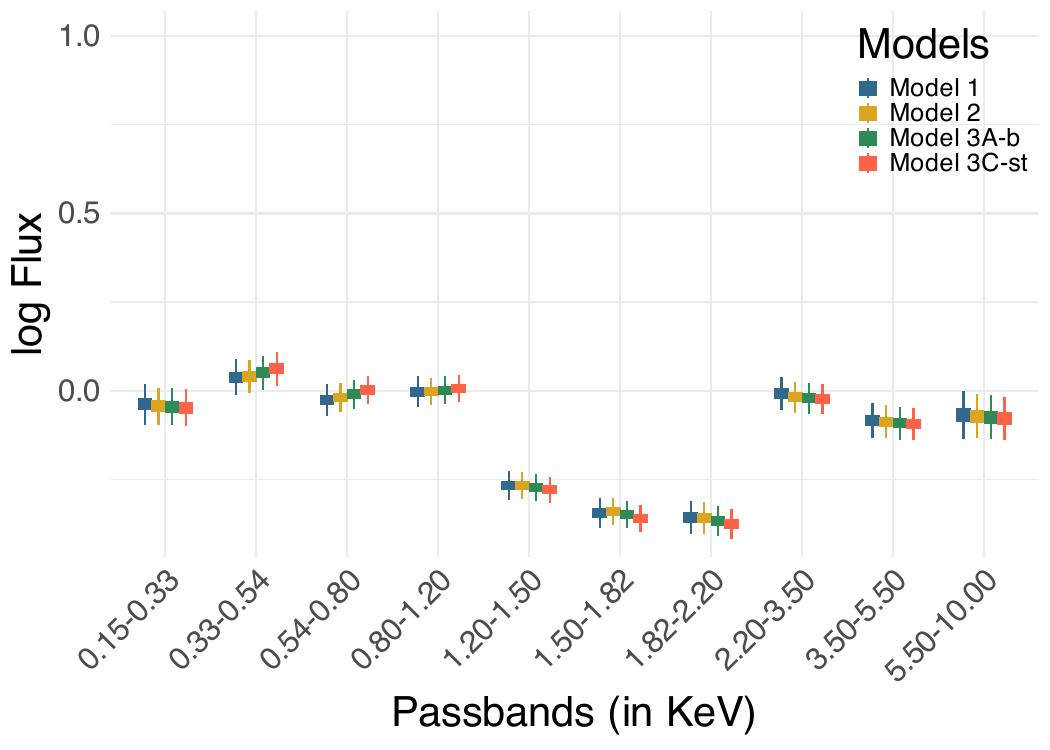}{0.33\textwidth}{(d) Source \texttt{6089}}\hfill
}
\gridline{
  \fig{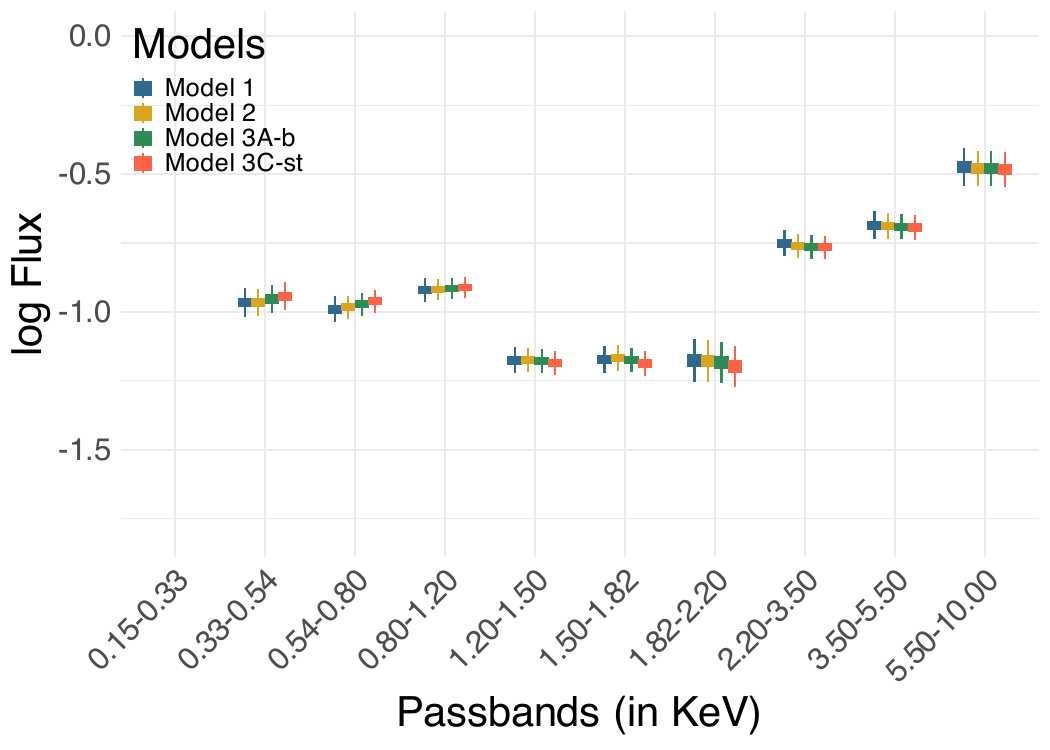}{0.33\textwidth}{(e) Source \texttt{4924B}}\hfill
  \fig{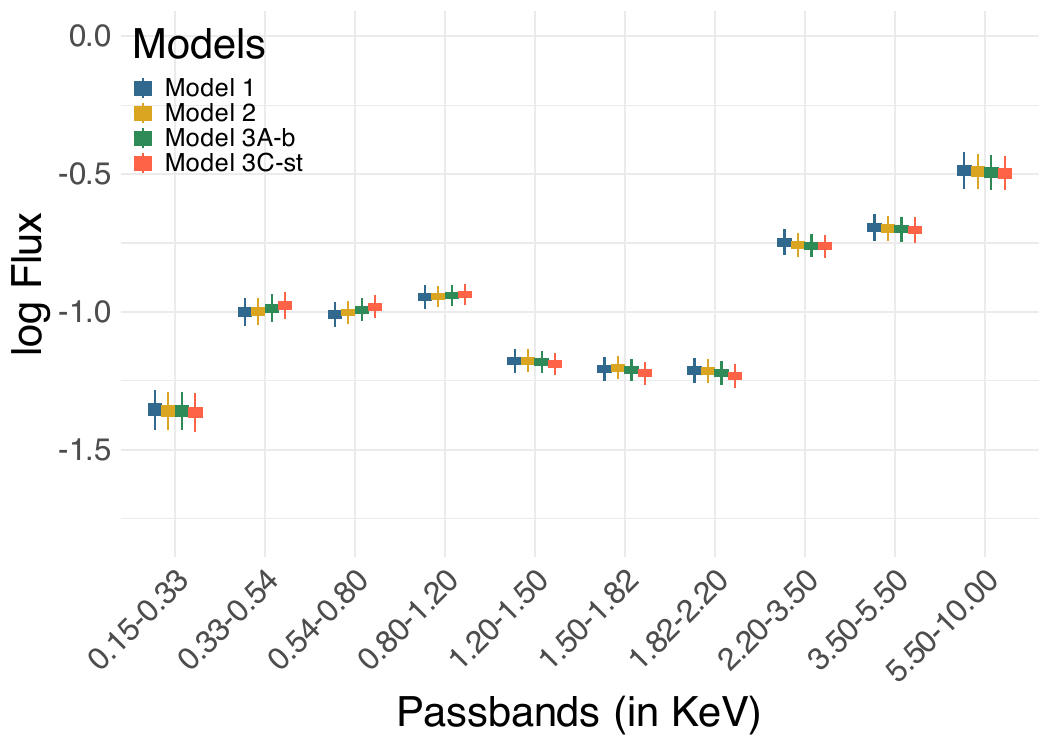}{0.33\textwidth}{(f) Source \texttt{4924C}}\hfill
}
\gridline{
  \fig{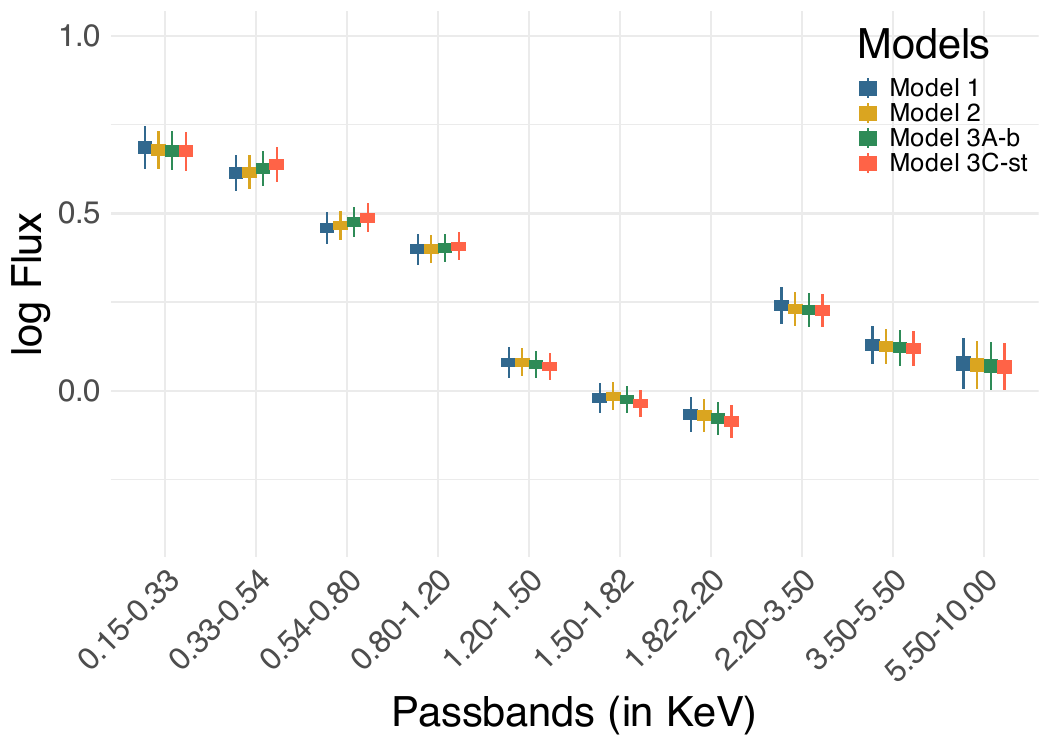}{0.33\textwidth}{(g) Source \texttt{1704}}\hfill
  \fig{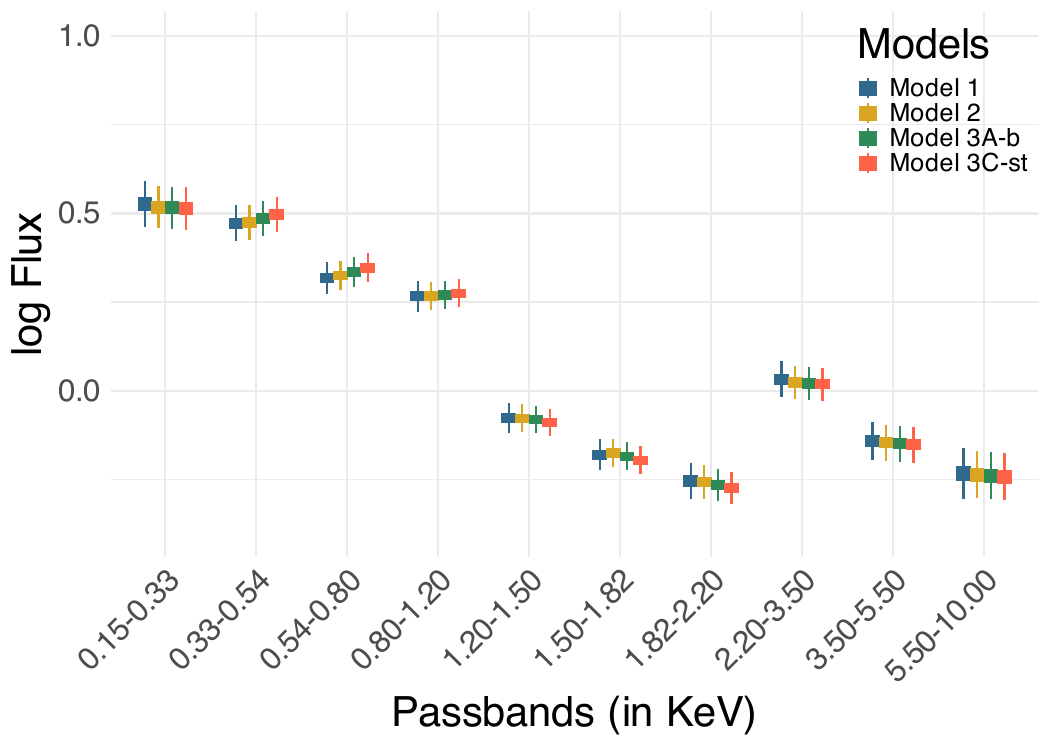}{0.33\textwidth}{(h) Source \texttt{4406}}\hfill
}
\caption{Boxplots of simulated posterior estimates of the log-flux estimate $\hat{G}_{sp}$ across ten passbands for eight sources under four models 1, 2, 3A-b and 3C-st. Boxplots from the same row correspond to same AGN target, but different sources. Since the distribution of photons over energy passbands inherently depends on the target, the plots of flux parameters coming from different sources but same target show similar pattern over energy passbands.}
\label{fig:appx_source_plot}
\end{figure*}

In Figure \ref{fig:appx_instrument_plot} and \ref{fig:appx_source_plot}, we have presented the boxplots of parameters $B_{ip}$ and $G_{sp}$ respectively. Figure \ref{fig:instrument_plot} and \ref{fig:appx_instrument_plot} jointly displays boxplots of effective area relative errors $B_{ip}$ for all instruments. For parameter $G_{sp}$, Figure \ref{fig:source_plot} and \ref{fig:appx_source_plot} jointly displays boxplots of log-flux parameter $G_{sp}$ for a total ten sources. 

%%%%%%%%%%%%%%%%%%%%%%%%%%%%%%%%%%%%%%%%%%%%%%%%%%%%%%%%%

\newpage
\section{Proofs}\label{appx:proofs}

%%%%%%%%%%%%%%%%%%%

\begin{proof}[Proof of Proposition \ref{prop:proper}]
Let $P_{is}$ be the set of passbands used to measure source $s$ by instrument $i$, $P_s$ be the set of passbands which includes observation for source $s$, $I_{sp}$ be the set of instruments employed to observe source $s$ in passband $p$, and $I_{s}$ be the set of instruments employed to observe source $s$. The cardinality of set $P_{is}$ is denoted as $|P_{is}|$, where $|P_s|,|P_{is}|\le P$, but $|P_s|\neq\sum_{i\in I_s}|P_{is}|$. The joint posterior density
\begin{align*}
    p\left(\boldsymbol{B}, \boldsymbol{G}, \boldsymbol{\sigma}^2 \mid \boldsymbol{y}\right) 
    &\propto p\left(\boldsymbol{y}\mid\boldsymbol{B}, \boldsymbol{G},\boldsymbol{\sigma}^2\right)p(\boldsymbol{B},\boldsymbol{G})p(\boldsymbol{\sigma}^2)\\
    &=\prod_{s}\prod_{i\in I_s}\prod_{p\in P_{is}} \frac{1}{\sqrt{\sigma_i^{2}}} \\
    &\hspace{0.6cm}\cdot\exp \left[-\frac{({y}_{isp} + \frac{\sigma_i^2}{2} - {B}_{ip} - {G}_{sp})^2}{2 \sigma_i^2}\right]\\
    &\hspace{0.3cm}\times\prod_{i} |\Gamma_i|^{-\frac{1}{2}} 
    \exp \left[-\boldsymbol{B}_{i}^\top \Gamma_i^{-1} \boldsymbol{B}_i \right] p(\sigma_i^2)\times\, p(\boldsymbol{G})
\end{align*}
where $p(\sigma_i^2)\propto\sigma_i^{-2\alpha-2}\exp\left[-\frac{\beta}{\sigma_i^2}\right]$. Models 1 and 2 assign non-informative prior, i.e., $p(\boldsymbol{G})\propto1$. According to our assumption, $|I_{sp}|\ge1$ for every $s$ and $p$. For a fixed source index $s$, if $\mathcal{I}$ is a random index such that $\text{Pr}(\mathcal{I} = i, \mathcal{I}\in I_{s}) \propto \sigma_i^{-2}$ and $\text{Pr}(\mathcal{I}\in I_s)=1$, then by Jensen's inequality
\begin{align*}
&\frac{\sum_{i \in I_{s}}\sigma_i^{-2}\sum_{p\in P_{is}} \left(y_{i s p}^{\prime}-B_{i p}-G_{s p}\right)^2}{\sum_{i \in I_{s}} \sigma_i^{-2}}\\
&\hspace{1.5cm}=\sum_{p=1}^P\frac{\sum_{i\in I_{sp}} \sigma_i^{-2}\left(y_{i s p}^{\prime}-B_{i p}-G_{s p}\right)^2}{\sum_{i \in I_{s}} \sigma_i^{-2}}\\
&\hspace{1.5cm}=\sum_{p=1}^{P} \mathbb{E}_{\mathcal{I}}\left[\left(y_{\mathcal{I} s p}^{\prime}-B_{\mathcal{I} p}-G_{s p}\right)\cdot\mathbf{1}\left(\mathcal{I}\in I_{sp}\right)\right]^2 \\
&\hspace{1.5cm}\geq\sum_{p=1}^{P} \left[\mathbb{E}_{\mathcal{I}}\left(\left({y}_{\mathcal{I}sp}^{\prime}-{B}_{\mathcal{I}p}-{G}_{sp}\right)\cdot\mathbf{1}\left(\mathcal{I}\in I_{sp}\right)\right)\right]^2\\
&\hspace{1.5cm}=\sum_{p=1}^{P} \left(\,\mathcal{C}_{sp}-G_{sp}\right)^2\cdot\text{Pr}(\mathcal{I}\in I_{sp})\\
&\hspace{1.5cm}=\sum_{p\in P_s}(\mathcal{C}_{sp} - {G}_{sp})^2\cdot\frac{\sum_{i\in I_{sp}}\sigma_i^{-2}}{\sum_{i\in I_s}\sigma_i^{-2}},
\end{align*}
where $y'_{isp}=y_{isp}+\frac{\sigma_i^2}{2}$. Here $\mathcal{C}_{sp}=\mathbb{E}_{\mathcal{I}}({y}_{\mathcal{I}sp}^\prime-{B}_{\mathcal{I}p})$ is independent of ${G}_{sp}$. Rewriting the inequality, we get
\begin{align*}
&\sum_{i \in I_{s}}\sum_{p\in P_{is}} \sigma_i^{-2}\left(y_{i s p}^{\prime}-B_{i p}-G_{s p}\right)^2\\
&\hspace{1.5cm}\ge\sum_{i\in I_{s}}\sum_{p\in P_{is}} (\mathcal{C}_{sp} - {G}_{sp})^2\sigma_i^{-2}.
\end{align*}
Thus, for some constant ${C}^*$,
\begin{align*}
    &p\left(\boldsymbol{B}, \boldsymbol{G}, \boldsymbol{\sigma}^2 \mid \boldsymbol{y}\right)\\
    &\hspace{1cm}\leq C^*\prod_i |\Gamma_i|^{-\frac{1}{2}} 
    \exp \left[-\boldsymbol{B}_{i}^\top \Gamma_i^{-1} \boldsymbol{B}_i \right] p(\sigma_i^2) \\
    &\hspace{2cm}\times \prod_{s}\left(\prod_{p\in P_{s}}\prod_{i\in I_{sp}}\sigma_i^{-1}\right)\\
    &\hspace{2.5cm}\cdot \exp \left[-\frac{1}{2} \left(\sum_{p\in P_s}(\mathcal{C}_{sp} - {G}_{sp})^2{\sum_{i\in I_{sp}}\sigma_i^{-2}}\right)\right].
\end{align*}
Integrating out $\boldsymbol{G}$ from this expression yields
\begin{align}\label{eqn:proof_prop_1}
    &\int p\left(\boldsymbol{B}, \boldsymbol{G}, \boldsymbol{\sigma}^2 \mid \boldsymbol{y}\right) \, d\boldsymbol{G}\nonumber\\
    &\leq C^{**} \prod_i |\Gamma_i|^{-\frac{1}{2}} 
    \exp \left[-\boldsymbol{B}_{i}^\top \Gamma_i^{-1} \boldsymbol{B}_i \right] p(\sigma_i^2) \nonumber\\
    &\hspace{1cm}\times \prod_{s}\prod_{p\in P_{s}}\left(\prod_{i\in I_{sp}}\sigma_i^{-1}\right)\left( \sum_{i\in I_{sp}} \sigma_i^{-2} \right)^{-1/2}.
\end{align}
Applying arithmetic mean-geometric mean inequality,
\begin{equation*}
    \left( \sum_{i\in I_{sp}} \sigma_i^{-2} \right)^{-1}
    \leq \frac{1}{|I_{sp}|} \left(\prod_{i\in I_{sp}}\sigma_i^{-2}\right)^{-1/|I_{sp}|}
    \leq \prod_{i\in I_{sp}}\left(\sigma_i^2\right)^{1/|I_{sp}|}.
\end{equation*}
We can always assume that $|I_{sp}|>0$, since if there is no observation in set $I_{sp}$ from source $s$ and passband $p$, then the corresponding parameter $G_{sp}$ is not estimable. Hence we can always assume $1\le |I_{sp}|\le I$ for all available combinations. We further integrate \eqref{eqn:proof_prop_1} over $\boldsymbol{B}$,
\begin{align*}
    &\int \int p(\boldsymbol{B},\boldsymbol{G},\boldsymbol{\sigma}^2\mid \boldsymbol{y})\,d\boldsymbol{G}\,d\boldsymbol{B} \\
    &\hspace{2cm}\leq C^{***}\prod_{s}\prod_{p\in P_{s}}\left(\prod_{i\in I_{sp}}\sigma_i^{-1+1/|I_{sp}|}\right) \prod_i p(\sigma_i^2)\\
    &\hspace{2cm}= C^{***} \prod_{i}\sigma_i^{-J_i}\left(\prod_{s\in S_{i}}\prod_{p\in P_{is}} \sigma_i^{\frac{1}{|I_{sp}|}}\right) \prod_i p(\sigma_i^2)\\
    &\hspace{2cm}= C^{***}  \prod_i \sigma_i^{-2\alpha-2-J_i+\sum_{s\in S_i}\sum_{p\in P_{is}}\frac{1}{|I_{sp}|}} e^{-\beta/\sigma_i^2}
\end{align*}
where $J_i$ is the total number of observations under instrument $i$.

If $\sigma_i \geq 1$, then 
\[J_i-\sum_{s\in S_i}\sum_{p\in P_{is}}\frac{1}{|I_{sp}|}=\sum_{s\in S_i}\sum_{p\in P_{is}}(1-\frac{1}{|I_{sp}|})\ge0,\]
and the marginal posterior can be bounded by an inverse-gamma density $\text{IG}(\alpha,\beta)$
\begin{align}\label{eqn:pf_prop_1}
    &\int \int p(\boldsymbol{B},\boldsymbol{G},\boldsymbol{\sigma}^2\mid \boldsymbol{y})\,d\boldsymbol{G}\,d\boldsymbol{B} \nonumber\\
    &\hspace{3cm}\leq C^{***}  \prod_i \sigma_i^{-2\alpha-2} e^{-\beta/\sigma_i^2}
\end{align}
which is finitely integrable with respect to $\boldsymbol{\sigma}^2$.

Alternatively, if $\sigma_i < 1$, then $\sigma_i^{-1+1/|I_{sp}|}\leq \sigma_i^{-1}$ and the exponent in \eqref{eqn:pf_prop_1} becomes $-(2\alpha+2+J_i)$; again giving an upper bound with inverse-gamma density $\text{IG}(\alpha+\frac{J_i}{2},\beta)$, which is also finitely integrable. Thus, the posterior is proper for Model 1 and 2.
\end{proof}

%%%%%%%%%%%%%%%%%%%

\begin{proof}[Proof of Proposition \ref{prop:identifiability_sigma}]
Posterior log-likelihood
\begin{align}\label{eqn:identifiability_sigma}
\log\,p(\boldsymbol{B},\boldsymbol{G},\boldsymbol{\sigma}^2\mid\boldsymbol{y}) &=\log\,p(\boldsymbol{y}\mid\boldsymbol{B},\boldsymbol{G},\boldsymbol{\sigma}^2)+\log\,p(\boldsymbol{B},\boldsymbol{G}\mid\boldsymbol{\sigma}^2)\notag\\
&\hspace{1.5cm}+\log\,p(\boldsymbol{\sigma}^2)+\text{ constant}
\end{align}
Lets define $r_{isp}=y_{isp}+\frac{\sigma_i^2}{2}-B_{ip}-G_{sp}$ such that $\frac{\partial r_{isp}}{\partial\sigma_i^2}=\frac{1}{2}$. Differentiating both sides in \eqref{eqn:identifiability_sigma} with respect to $\sigma_i^2$ we get
\begin{align*}
    \frac{\partial}{\partial \sigma_i^2} \log p(\boldsymbol{y}\mid \boldsymbol{B},\boldsymbol{G},\boldsymbol{\sigma}^2)
    &=\frac{\partial}{\partial \sigma_i^2} \left[ -\frac12 \sum_{s,p} \left( \log(2\pi \sigma_i^2) + \frac{r_{isp}^2}{\sigma_i^2}\right)\right]\notag\\
    &=-\frac12 \sum_{s,p}
    \left[\frac{1}{\sigma_i^2}+\frac{\partial}{\partial \sigma_i^2}\!\left(\frac{r_{isp}^2}{\sigma_i^2}\right)\right]\notag\\
    &=-\frac12 \sum_{s,p}\left[\frac{1}{\sigma_i^2}+\frac{r_{isp}}{\sigma_i^2}-\frac{r_{isp}^2}{(\sigma_i^2)^2}\right].
\end{align*}
Since $\sigma_i^2$ has prior $\text{Inv-Gamma}(\alpha,\beta)$ and the priors of $\boldsymbol{B}$ and $\boldsymbol{G}$ does not depend on $\boldsymbol{\sigma}$, then
\begin{align*}
\frac{\partial}{\partial \sigma_i^2}
\log p(\boldsymbol{B},\boldsymbol{G}\mid \boldsymbol{\sigma}^2)&=0
\end{align*}
and
\begin{align*}
\frac{\partial}{\partial\sigma_i^2}\log p(\sigma_i^2)
=-\frac{\alpha+1}{\sigma_i^2}+\frac{\beta}{\sigma_i^4}.
\end{align*}

The first derivative of \eqref{eqn:identifiability_sigma} will be
\begin{align}\label{eqn:posterior_maxima}
\frac{\partial}{\partial \sigma_i^2}\ell(\boldsymbol{B},\boldsymbol{G},\boldsymbol{\sigma}^2)
&=-\frac12 \sum_{s,p}\left[\frac{1}{\sigma_i^2}+\frac{r_{isp}}{\sigma_i^2}-\frac{r_{isp}^2}{\sigma_i^4}\right]-\frac{\alpha+1}{\sigma_i^2}+\frac{\beta}{\sigma_i^4}.
\end{align}
Note that the derivative in \eqref{eqn:posterior_maxima} depends only on $\sigma_i$. Equating it to zero we get,
\begin{equation}\label{eqn:identifiability_maxima}
\frac12 \sum_{s,p}\left[\frac{1}{\sigma_i^2}+\frac{r_{isp}}{\sigma_i^2}-\frac{r_{isp}^2}{\sigma_i^4}\right]=-\frac{\alpha+1}{\sigma_i^2}+\frac{\beta}{\sigma_i^4}.
\end{equation}
We can easily show that \eqref{eqn:identifiability_maxima} is a quadratic equation of $\sigma_i^{-2}$ with a real-valued solution. If we have a posterior maxima at $\widehat{\boldsymbol{\sigma}}^2$, it should satisfy the identity in \eqref{eqn:identifiability_maxima} for every instrument $i$, and the second derivative at $\widehat{\sigma}_i^2$ should be negative.
\begin{align*}
&\frac{\partial^2}{\partial (\sigma_i^2)^2}
\ell(\boldsymbol{B},\boldsymbol{G},\boldsymbol{\sigma}^2)\\
&\hspace{1cm}=-\frac12 \sum_{s,p}
\left[-\frac{1}{\sigma_i^4}+\frac{1}{2\sigma_i^2}-\frac{2r_{isp}}{\sigma_i^4}+\frac{2r_{isp}^2}{\sigma_i^6}\right]
+\frac{\alpha+1}{\sigma_i^4}-\frac{2\beta}{\sigma_i^6}\\
&\hspace{1cm}=-\frac12 \sum_{s,p}
\left[-\frac{1}{\sigma_i^4}+\frac{1}{2\sigma_i^2}-\frac{2r_{isp}}{\sigma_i^4}+\frac{2r_{isp}^2}{\sigma_i^6}\right]\\
&\hspace{4cm}-\frac{1}{2\sigma_i^2} \sum_{s,p}\left[\frac{1}{\sigma_i^2}+\frac{r_{isp}}{\sigma_i^2}-\frac{r_{isp}^2}{\sigma_i^4}\right]-\frac{\beta}{\sigma_i^6}\\
&\hspace{1cm}=-\frac12 \sum_{s,p}
\left[\frac{1}{2\sigma_i^2}-\frac{r_{isp}}{\sigma_i^4}+\frac{r_{isp}^2}{\sigma_i^6}\right]-\frac{\beta}{\sigma_i^6}\\
&\hspace{1cm}=-\frac{1}{2\sigma_i^2}\sum_{s\in S_i,\,p\in P_{is}} \left[\left(\frac{r_{isp}}{\sigma_i^2}-1\right)^2 + \left(\frac{r_{isp}}{\sigma_i^2}\right)^2\right]-\frac{\beta}{\sigma_i^6}\quad<\,\,0\,.
\end{align*}
This proves that the $\boldsymbol{\sigma}^2$ parameter can be uniquely estimated from the posterior distribution given the observations $\boldsymbol{y}$.
\end{proof}

%%%%%%%%%%%%%%%%%%%

\begin{proof}[Proof of Proposition \ref{prop:identifiability}]
The conditional posterior distribution is given by 
\begin{align*}
&p\left(\boldsymbol{B}, \boldsymbol{G} \mid \boldsymbol{y}, \boldsymbol{\sigma}^2\right)\\
&\propto p\left(\boldsymbol{y} \mid \boldsymbol{B}, \boldsymbol{G}, \boldsymbol{\sigma}^2\right) 
        \, p\left(\boldsymbol{B},
        \boldsymbol{G} \mid \boldsymbol{\sigma}^2\right) \\
&\propto \prod_{i,s,p} \exp\left[ 
        -\frac{\left(y_{isp} + \frac{\sigma_i^2}{2} - B_{ip} - G_{sp} \right)^2}{2\sigma_i^2} 
        \right] \\
&\hspace{2cm} \times \prod_i \exp\left( 
        -\frac{1}{2} \boldsymbol{B}_i^\top \Gamma_i^{-1} \boldsymbol{B}_i 
        \right),
\end{align*}
where matrices $\Lambda$, $\Delta$, $\Theta$, $\Sigma$ and vectors $\boldsymbol{K}$, $\boldsymbol{Q}$ are defined in the notations below. Thus, the joint posterior of $(\boldsymbol{B},\boldsymbol{G})$ is multivariate normal
\[
\begin{bmatrix} \boldsymbol{B} \\ \boldsymbol{G} \end{bmatrix} \mid \boldsymbol{y},\boldsymbol{\sigma}^2 \sim \mathcal{N}(\boldsymbol{\mu}_{BG}, \Sigma_{BG}),
\]
where
\[
\boldsymbol{\mu}_{BG} = \begin{bmatrix} \boldsymbol{\mu}_B \\ \boldsymbol{\mu}_G \end{bmatrix}, \quad \Sigma_{BG} = \begin{bmatrix} \Lambda^{-1} & \Theta \\ \Theta^\top & \Delta^{-1} \end{bmatrix}^{-1}.
\]
The posterior means satisfy equations
\begin{align*}
\Lambda^{-1} \boldsymbol{\mu}_B + \Theta \boldsymbol{\mu}_G &= \Sigma^{-1} \boldsymbol{K},\\
\Delta^{-1} \boldsymbol{\mu}_G + \Theta^\top \boldsymbol{\mu}_B &= \boldsymbol{Q},
\end{align*}
which can be explicitly solved as
\begin{align*}
\boldsymbol{\mu}_B &= \left(\Lambda^{-1} - \Theta \Delta \Theta^\top\right)^{-1}\left( \Sigma^{-1}\boldsymbol{K} - \Theta\Delta\boldsymbol{Q} \right),\\
\boldsymbol{\mu}_G &= \Delta\boldsymbol{Q} - \Delta\Theta^\top\boldsymbol{\mu}_B.
\end{align*}
The symbols and notations employed above are detailed as follows.
\begin{align*}
\boldsymbol{B}_i &= (B_{i1}, \ldots, B_{iP})^\top, 
\hspace{-0.4cm} &
\boldsymbol{G}_s &= (G_{s1}, \ldots, G_{sP})^\top, \\
\Sigma_i &= \sigma_i^2 I_P, 
\hspace{-0.4cm} &
\Sigma &= \operatorname{diag}(\Sigma_1, \ldots, \Sigma_I), \\
\Theta &= \boldsymbol{1}_S^\top \otimes (\Sigma_1^{-1}, \ldots, \Sigma_I^{-1})^\top, 
\hspace{-0.4cm} &
\Lambda_i^{-1} &= S \Sigma_i^{-1} + \Gamma_i^{-1}, \\
\Lambda &= \operatorname{diag}(\Lambda_1, \ldots, \Lambda_I), 
\hspace{-0.4cm} &
\Delta^{-1} &= I_S \otimes \left( \sum_i \Sigma_i^{-1} \right), \\
\boldsymbol{K}_i &= \sum_s \left( \boldsymbol{y}_{is} + \frac{\sigma_i^2}{2} \boldsymbol{1}_P \right), 
\hspace{-0.4cm} &
\boldsymbol{K} &= (\boldsymbol{K}_1, \ldots, \boldsymbol{K}_I)^\top, \\
\boldsymbol{Q}_s &= \sum_i \Sigma_i^{-1} \left( \boldsymbol{y}_{is} + \frac{\sigma_i^2}{2} \boldsymbol{1}_P \right), 
\hspace{-0.4cm} &
\boldsymbol{Q} &= (\boldsymbol{Q}_1, \ldots, \boldsymbol{Q}_S)^\top.
\end{align*}
To ensure identifiability, it is necessary to show that $\Sigma_{BG}^{-1}$ is positive definite. By Schur complement condition, $\Sigma_{BG}^{-1}$ is positive definite if and only if
(a) $\Lambda$ is positive definite (holds by construction), and
(b) $\Lambda^{-1} - \Theta\Delta\Theta^\top$ is positive definite.

Direct calculation yields
\[\Theta\Delta\Theta^\top = S \left( \boldsymbol{\eta} \boldsymbol{\eta}^\top / (\boldsymbol{\eta}^\top \boldsymbol{1}) \right) \otimes I_P,\]
where $\boldsymbol{\eta} = (\sigma_1^{-2},\ldots,\sigma_I^{-2})^\top$. Then
\begin{align*}
    \Lambda^{-1} - \Theta\Delta\Theta^\top &= S\left( \operatorname{diag}(\boldsymbol{\eta}) - \frac{\boldsymbol{\eta}\boldsymbol{\eta}^\top}{\boldsymbol{\eta}^\top\boldsymbol{1}} \right) \otimes I_P + (\operatorname{diag}(\Gamma_i)_{i=1}^I)^{-1}.
\end{align*}
The matrix $\left(\operatorname{diag}(\boldsymbol{\eta}) - \frac{\boldsymbol{\eta}\boldsymbol{\eta}^\top}{\boldsymbol{\eta}^\top\boldsymbol{1}}\right)$ is non-negative definite by the Cauchy-Schwarz inequality. Since $\Gamma_i$'s are positive definite, $\operatorname{diag}(\Gamma_i^{-1})$ is also positive definite. Thus, $\Lambda^{-1} - \Theta\Delta\Theta^\top$ is a positive definite matrix, completing the proof.
\end{proof}

%%%%%%%%%%%%%%%%%%%%%%%%%%%%%%%%%%%%%%%%%%%%%%%%%%

\section{Missing Value Estimation And Predictive p-values}\label{sec:missing_estimation}\label{appx:pvalue}

In this section, we can conduct a posterior predictive performance evaluation to further assess the robustness of our model. After imposing prior constraints on parameters of the initially weakly identifiable model, resulting we get a hierarchical Bayesian model setup to estimate the effective area, source flux, and instrument-specific variance parameters. The posterior inference may still be sensitive to prior specifications, modeling assumptions, and imbalance in the observed data. The uneven observation frequencies across instruments, as reported in Table~\ref{tab:instrument_counts}, may lead to less stable estimation for instruments with relatively sparse data. To assess the robustness of the posterior inference, we perform a sensitivity analysis based on posterior predictive p-values and leave-one-out predictive p-values.

For every existing data point $y_{isp}$, the posterior predictive distribution is given by
\begin{equation*}
    \hat{y}_{isp} \sim \mathcal{N}\left(-\frac{\hat{\sigma}_i^2}{2} + \hat{B}_{ip} + \hat{G}_{sp},\,\,\hat{\sigma}_i^2\right),
\end{equation*}
where $\hat{B}_{ip},\,\hat{G}_{sp},\,\hat{\sigma}_i^2$ are the posterior samples of corresponding model parameters obtained using Hamiltonian Monte Carlo. We calculate the posterior predictive p-value for each of our available data $y_{isp}$, by comparing the posterior predictive samples $\{\hat{y}_{isp}^1,\ldots,\hat{y}_{isp}^K\}$ against the observed value of $y_{isp}$,
\begin{equation*}
\mathrm{p}_{(isp)}^{\text{full}}=2\times\min\left\{\frac{1}{K}\sum_{k=1}^K\mathbf{1}\left(\hat{y}_{isp}^k<y_{isp}\right),\frac{1}{K}\sum_{k=1}^K\mathbf{1}\left(\hat{y}_{isp}^k>y_{isp}\right)\right\}
\end{equation*}
We get one predictive p-value from each of the 2098 observed data $y_{isp}$. Only 3.1\% of the predictive p-values are less than $0.05$, which indicates satisfactory overall predictive fit. Moreover, these small p-values are not concentrated among low-frequency instruments, but instead occur across instruments, suggesting that lack of fit is not driven solely by sparsely observed instrument groups.

Recall that our model parameters effective area relative error $B_{ip}$ and log-flux $G_{sp}$ depend on joint instrument-passband indices and on joint source-passband indices respectively. In order to predict a particular data, we must have at least one observation from the same instrument-passand index combination (table \ref{tab:instrument_counts}) and one observation from the same source-passband index combination (table \ref{tab:source_counts}). In other words, if we want to predict the value $y_{isp}$ using our fitted model, then we need at least one observation from each of the sets $\{y_{is^\prime p}\}_{s^\prime\ne s}$ and $\{y_{i^\prime sp}\}_{i^\prime\ne i}$. If this condition is satisfied, we can estimate the missing observation as
\begin{equation*}
    \hat{y}_{isp}=-\frac{1}{2}\hat{\sigma}_i^2+\hat{B}_{ip}+\hat{G}_{sp}.
\end{equation*}

We take advantage of this estimation framework to obtain leave-one-out p-values, where we predict $y_{isp}$ based on data $\mathbf{y}$, excluding the data point itself. We carefully remove the log-photon data $y_{isp}$, ensuring that the frequencies in Tables \ref{tab:source_counts} and \ref{tab:instrument_counts} do not drop to zero any further than there already are. We compare the observed ${y}_{isp}$ with the leave-one-out predictive samples $\hat{y}_{isp}^{\text{loo}}$ to get the LOO p-values $\mathrm{p}_{isp}^{\text{loo}}$. We can calculate such leave-one-out (LOO) predictive p-values for 2083 out of all 2098 many available data points that satisfy the frequency table condition.

Since the full-data predictive distribution uses the observation $y_{isp}$ when fitting the model, whereas the leave-one-out predictive distribution does not, we generally expect $\mathrm{p}_{isp}^{\mathrm{loo}}$ to be smaller than $\mathrm{p}_{isp}^{\mathrm{full}}$. Only 4.9\% out of 2083 LOO predictive p-values are less than 0.05. This implies that even under the absence of the true value of the data, our model can correctly predict the data $y_{isp}$.

%%%%%%%%%%%%%%%%%%%%%%%%%%%%%%%%%%%%%%%%%%%%%%%%%%

%%%%%%%%%%%%%%%%%%%%%%%%%%%%%%%%%%%%%%%%%%%%%%%%%%
%%%%%%%%%%%%%%%%%%%%%%%%%%%%%%%%%%%%%%%%%%%%%%%%%%
%%%%%%%%%%%%%%%%%%%%%%%%%%%%%%%%%%%%%%%%%%%%%%%%%%



\bibliographystyle{rasti}
\bibliography{references} % if your bibtex file is called example.bib


%If you want to present additional material which would interrupt the flow of the main paper, it can be placed in an Appendix which appears after the list of references.

%%%%%%%%%%%%%%%%%%%%%%%%%%%%%%%%%%%%%%%%%%%%%%%%%%

% Don't change these lines
\bsp	% typesetting comment
\label{lastpage}
\end{document}